\documentclass[12pt]{article}
\usepackage[utf8]{inputenc}
\usepackage[export]{adjustbox}
\usepackage{authblk}
\usepackage{bbm}
\usepackage{pgfpages}
\usepackage{graphicx}
\usepackage{multicol}
\usepackage{csquotes}
\usepackage{multirow}
\usepackage{graphicx} % Allows including images
\usepackage{booktabs} % Allows the use of \toprule, \midrule and \bottomrule in table    s
\usepackage{authblk}
\usepackage{caption}
\usepackage{subcaption}
\usepackage{stackengine}
\usepackage{amsmath}
\usepackage{comment}
\usepackage{hyperref}
\hypersetup{
    colorlinks=true,
    linkcolor=blue,
    filecolor=blue,      
    urlcolor=blue,
    citecolor=blue,
}
\usepackage[capposition=top]{floatrow}

\usepackage[spanish,english]{babel}

\usepackage{amssymb}
\usepackage{stackrel}
\usepackage{natbib} % bibliography

\usepackage[a4paper, margin=1in]{geometry}

\title{The Transmission of US Monetary Policy Shocks \\ \Large The Role of Investment \& Financial Heterogeneity}
\author[1]{Santiago Camara\footnote{Address: 2211 Campus Dr, Evanston, IL 60208. E-mail: santiagocamara2022@u.northwestern.edu . I would like to thank the Banco Central de Chile for financial support. Also, I would like to thank Marty Eichenbaum, Lawrence Christiano, Giorgio Primiceri and Matthew Rognlie who provided important guidance. Finally, I would also like to thank Javier Garcia Cicco and Mauricio Calani for their comments. }}
\author[2]{Sebastian Ramirez Venegas}
\affil[1]{Northwestern University \& Red-NIE}
\affil[2]{CMF (Comisi\'on para el Mercado Financiero)}

\date{\today}

\begin{document}
    
\maketitle

\begin{abstract}
%\large
This paper studies the transmission of US monetary policy shocks into Emerging Markets emphasizing the role of investment and financial heterogeneity. First, we use a panel SVAR model to show that a US interest tightening leads to a persistent recession in Emerging Markets driven by a sharp reduction in aggregate investment. Second, we study the role of firms' financial heterogeneity in the transmission of US interest rate shocks by exploiting detailed balance sheet dataset from Chile. We find that more indebted firms experience greater drops in investment in response to a US tightening shock than less indebted firms. This result is at odds with recent evidence from US firms, even when using the same identification strategy and econometric methods. Third, we rationalize this finding using a stylized model of heterogeneous firms subject to a tightening leverage constraint. Finally, we present evidence in support of this hypothesis as well as robustness checks to our main results. Overall, our results suggests that the transmission channel of US monetary policy shocks within and outside the US differ, a result novel to the literature.

\medskip

\medskip


\medskip

\end{abstract}

%%%%%%%%%%%%%%%%%%%%%%%%%%%%%%%%%%%%%%%%%%%%%%%%%%%%%%%%%%%%%%%%%%%%%%
%% Introduction
\newpage
\section{Introduction} \label{sec:introduction}

The impact of US monetary policy on Emerging Market  (EM) economies' is a classic question in international macroeconomics, going back to \cite{fleming1962domestic}, \cite{mundell1963capital}, \cite{dornbusch1976expectations} and \cite{frenkel1983monetary}. The status of the US dollar as the global reserve currency, unit of account, invoice of international trade, its dominant role in financial markets and the global financial cycles implies that the Federal Open Market Committee's (FOMC) policy decisions have spillover effects on the rest of the world. While a recent literature finds that US monetary policy shocks leads to significant recessions in EMs, the transmission channels these shocks is still an open question (see \cite{degasperi2020global,camara2021spillovers}). In this paper, we use aggregate data to show that US interest rates primarily affect EM economies through a persistent fall in investment. Additionally, using detailed micro level data from Chile we show that this fall in investment is driven by firms facing tighter borrowing constraints. These results are at odds with a recent literature which study the response of US firms to monetary policy shocks.  

First, we address the question on the aggregate transmission of US monetary policy shocks to EM economies through panel SVAR models. Our empirical approach combines US monetary policy shocks, measured using orthogonalized high frequency event-study approach sourced from \cite{bauer2022reassessment}, with macroeconomic and financial variables from 8 EM economies. We find that a US monetary policy shock leads to a persistent drop in economic activity with a fall in investment explaining between half and two-thirds of it. We show that this result is robust across different variable and sample specifications.

Second, we use firm level data from Chile to address the question on the micro level transmission of US monetary policy shocks through lower aggregate investment. Our empirical approach combines the US monetary policy shocks introduced in the panel SVAR model, with quarterly balance sheet data from Chile's \textit{Comision para el Mercado Financiero} or CMF. We find that investment done by more indebted firms is significantly and persistently more responsive to US monetary policy shocks than investment done by relatively less indebted firms. This result is in stark contrast with results found for US firms by \cite{ottonello2020financial}, even when using the same empirical approach. In particular, our baseline empirical specification estimates how the semi-elasticity of firm investment with respect to a US monetary policy shock depends on firms' leverage ratio. We control for firm fixed effects to capture permanent differences across firms and control for sector-by-quarter fixed effects to capture differences in how sectors respond to aggregate shocks. 

We find that having one standard deviation higher leverage implies that a firm is approximately one-fourth more responsive to monetary policy and that having one standard deviation higher distance to default implies that the firm is one-half more responsive. These differences across firms persist for up to two years after the shock and imply large differences
in accumulated capital over time. We show that our results are robust to different econometric specifications which characterize firms' heterogeneity in their financial position. Additionally, we show that these result are robust to measures of both short and long term leverage and when controlling for firms' liability currency mismatch. %Furthermore, we show that heterogeneity in firms' currency mismatch does not consistently predict differences in investment patterns in response to a US monetary policy shock.

Third, motivated by our micro level evidence we construct two stylized models of heterogeneous entrepreneurs to show how different transmission channels affect their investment. On the one hand, we show that an increase of borrowing interest rates predict a greater response of the investment of firms relatively less indebted compared to relatively more indebted firms, which are relatively closer to their borrowing constraint. On the other hand, we show that a tighter borrowing constraint, predict a greater response of the investment of firms relatively more indebted while either not affecting or even increasing the investment of less indebted firms. Consequently, we interpret our micro level evidence as suggesting that the transmission of US monetary policy shocks into EM economies is primarily driven by tighter borrowing constraints rather than by higher borrowing rates.

%Lastly, we provide supporting evidence of the hypothesis of US monetary policy shocks triggering tighter borrowing conditions in EM. To this end, we follow a two prong approach. First, we augment our SVAR model using additional variables. We show that US monetary policy shocks does not impact significantly on domestic borrowing rates but lead to lower domestic lending and significant capital out-flows. Second, we show that 

\noindent
\textbf{Related literature.} This paper relates to three main strands of literature. First, this paper contributes to a long strand of literature which has focused on identifying and quantifying the impact of US monetary policy shocks in small open economies. Early examples of this literature are \cite{eichenbaum1995some} and \cite{uribe2006country} using data from the 1980s, 1990s and early 2000s; and more recently \cite{vicondoa2019monetary} and \cite{camara2021spillovers} using data up until the late 2010s. The vast majority of this literature finds that US monetary policy shocks lead to an economic recession, and exchange rate depreciation and overall tighter financial conditions. However, a recent literature focusing on evidence for the last two decades, has found atypical dynamics, in which US interest rate shocks have been associated with economic expansions. Our contribution to this literature is finding evidence consistent with a recessionary effect of US monetary policy shocks by using an identification strategy that cleanses any potential bias that the systematic disclosure of information around of FOMC meetings may introduce.\footnote{This result is in line with the results presented by \cite{camara2021spillovers}. However, while the identification strategy used in \cite{camara2021spillovers} jointly identifies two FOMC shocks, a pure monetary policy shock and an information disclosure shock, in the present paper we rely on the identification strategy introduced by \cite{bauer2022reassessment} which specifically control for any possible bias introduced by the Federal Reserve's response to news channel. Thus, our benchmark aggregate results complement those presented by \cite{camara2021spillovers}. } Consequently, this paper shows that US monetary policy shocks are associated with negative spillovers on EM economies, even for recent samples. 

Second, this paper relates to the study of the transmission channels of US monetary policy shocks on the rest of the world. Early examples of this literature are 
\cite{akram2009commodity} and \cite{dedola2017if}. However, \cite{degasperi2020global} is particularly close to our paper. The authors use a large scale SVAR model with more than 30 variables from both the US and panel of Advanced and EM economies to study the aggregate transmission of US monetary policy shocks. In particular, the authors find that financial channels dominate over demand and exchange rate channels in the transmission to real variables, while the transmission
via commodity prices determines nominal spillovers. Our aggregate results are in line with their findings as we find that US monetary policy shocks lead to tighter financial conditions associated with a significant drop in aggregate investment. We build on this aggregate findings by using firm level data. In particular, we exploit firms' financial heterogeneity to gauge the relative importance of higher borrowing rates and tighter financial conditions to disentangle the micro level transmission of US monetary policy shocks into lower aggregate investment. 

Third, this paper relates to a relatively new literature which exploits micro level data to study the transmission of US interest rate shocks in small open economies. For instance, 
\cite{camara2022tank} and \cite{auclert2021exchange} study how household heterogeneity matter for the transmission of exchange rate shocks, proxied through foreign interest rate shocks, using a calibrated heterogeneous agent New Keynesian models. We differ by focusing on the role of firms' financial heterogeneity in the transmission of US monetary policy shocks. Thus, our paper is close to \cite{ottonello2020financial} which studies the same question but for US firms. Interestingly, our results are completely at odds with those found by \cite{ottonello2020financial}, even when implementing the same econometric specifications and identification strategy. In particular, \cite{ottonello2020financial} finds that relatively less indebted firms are more responsive to US monetary policy shocks in contrast to relatively more indebted firms. We argue that this suggests that the transmission channels of US monetary policy shocks within and outside US are different, a result novel to the literature of international economics.

\noindent
\textbf{Organization.} This paper is comprised of \ref{sec:conclusion} sections starting with the current introduction. Section \ref{sec:aggregate_evidence} presents our results on the aggregate transmission of US monetary policy shocks for a panel of EM economies. Section \ref{sec:firm_evidence} presents our micro level results by showing that firms' heterogeneous financial position matters for their investment response to US monetary policy shocks, finding that relatively more indebted firms are more responsive than relatively less indebted. In Section \ref{sec:stylized_model} we interpret our firm level results through two stylized structural models of tighter borrowing constraints. Section \ref{sec:conclusion} concludes.  %Section \ref{sec:supporting_evidence} presents both aggregate and firm level supporting evidence. 

%%%%%%%%%%%%%%%%%%%%%%%%%%%%%%%%%%%%%%%%%%%%%%%%%%%%%%%%%%%%%%%%%%%%%%
%% Aggregate Evidence
\section{Aggregate Evidence} \label{sec:aggregate_evidence}

In this section of the paper we provide aggregate evidence that US interest rate shocks lead to a persistent reduction in economic output primarily driven by a drop in aggregate investment. In Section \ref{subsec:aggregate_evidence_panel} we find supporting evidence of this hypothesis by estimating a panel SVAR for a panel of Emerging Market economies. In Section \ref{subsec:aggregate_evidence_chile} we show that this hypothesis also holds for the Chilean economy, which is the Emerging Market economy for which we count with detailed firm-level data used in Section \ref{sec:firm_evidence}.

\subsection{Evidence for a Panel of Emerging Market Economies} \label{subsec:aggregate_evidence_panel}

We start by describing the SVAR model estimated in this section of the paper. The model specification is a pooled panel SVAR as presented by \cite{canova2013panel}. This type of model considers the dynamics of several countries simultaneously, but assuming that the dynamic coefficients are homogeneous across units, and coefficients are time-invariant. In this framework, this implies that Emerging Market country $i$'s variables only depend on US variables and the lagged values of country $i$'s variables. Although the possible interactions and inter-dependencies across Emerging Markets is an interesting topic on itself, we abstract from this considerations in this paper. 

In its most general form, a panel SVAR model comprises of N countries or units, $n$ endogenous variables, $p$ lagged values and $T$ time periods. The pooled panel SVAR model can be written as
\begin{align} \label{eq:pooled_estimator}
\begin{pmatrix}
y_{1,t} \\
y_{2,t} \\
\vdots  \\
y_{N,t}
\end{pmatrix}
&=C+
\begin{pmatrix}
A^1 \quad 0 \quad \cdots \quad 0 \\
0 \quad  A^1 \quad \cdots \quad 0 \\
\vdots \quad \vdots \quad \ddots \quad \vdots \\
0 \quad 0 \quad \cdots \quad A^1 
\end{pmatrix}
\begin{pmatrix}
y_{1,t-1} \\
y_{2,t-1} \\
\vdots  \\
y_{N,t-1}
\end{pmatrix}
+ \cdots \nonumber \\
\\
&+ \nonumber
\begin{pmatrix}
A^p \quad 0 \quad \cdots \quad 0 \\
0 \quad  A^p \quad \cdots \quad 0 \\
\vdots \quad \vdots \quad \ddots \quad \vdots \\
0 \quad 0 \quad \cdots \quad A^p 
\end{pmatrix}
\begin{pmatrix}
y_{1,t-p} \\
y_{2,t-p} \\
\vdots  \\
y_{N,t-p}
\end{pmatrix}
+
%\begin{pmatrix}
%C \\
%C \\
%\vdots \\
%C
%\end{pmatrix}
%x_t + 
\begin{pmatrix}
\epsilon_{1,t} \\
\epsilon_{2,t} \\
\vdots \\
\epsilon_{N,t}
\end{pmatrix}
\end{align}
where $y_{i,t}$ denotes an $n \times 1$ vector of $n$ endogenous variables of country $i$ at time $t$ and $A^{j}$ is an $n \times n$ matrix of coefficients providing the response of country $i$ to the $j^{th}$ lag at period $t$. Note that by assuming that $A^j_1 = A^j_n = A^j$ for $j=1,\ldots,n$ implies the assumption that the estimated coefficients are common across countries. $C$ is a $Nn\times1$ vector of constant terms which are also assumed to be common across countries. Lastly, $\epsilon_{i,t}$ is an $n \times 1$ vector of residuals for the variables of country $i$, such that
\begin{align*}
    \epsilon_{i,t} \sim \mathcal{N}\left(0,\Sigma_{ii,t}\right)
\end{align*}
with 
\begin{align*}
    \epsilon_{ii,t} &= \mathbb{E} \left(\epsilon_{i,t} \epsilon_{i,t}' \right) = \Sigma_c \quad \forall i \\
    \epsilon_{ij,t} &= \mathbb{E} \left(\epsilon_{i,t} \epsilon_{j,t}' \right) = 0 \quad \text{for } i \neq j
\end{align*}
The last two equations imply that, as for the model's auto-regressive coefficients, the innovation's variance is equal across countries. 

Next, we describe the different type of variables that comprise vector $y_t$. This vector is comprised of three types of variables: $m_t$ which denotes identified exogenous US monetary policy shocks; $y^{*}_t$ which denotes US variables, and $\tilde{y}_t$ which denotes the vector of Emerging Market variables. In terms of the vector of high-frequency variables $m_t$, our benchmark empirical specification uses the orthogonalized high-frequency shocks constructed by \cite{bauer2022reassessment}.\footnote{\cite{bauer2022reassessment} construct these shocks by studying the high-frequency surprises of interest rates around both FOMC meetings and speeches by the Chairman of the Federal Reserve. In 2.6\% of the months in our sample have more than one high-frequency surprise. To deal with these multiple shocks we take median shock. Other specifications, such as taking the mean of the shocks within the month, or outright excluding the months with multiple observations do not alter the results significantly or quantitatively.} The vector of US variables $y^{*}_t$ is comprised of the 10 year US treasury constant maturity rate, the corporate spread spread constructed by \cite{gilchrist2012credit}, real GDP and the PCE consumer price index. This specification is in line with the literature who studies the impact of US interest rate shocks in the US economy (see \cite{jarocinski2020deconstructing,bauer2022reassessment}). In Appendix \ref{sec:appendix_model_svar_results} we show that results are robust to different specifications of US variables. Vector $\tilde{y}_t$ is comprised of EM's EMBI spreads, monetary policy rate, consumer price index, nominal exchange rate, interpolated GDP and aggregate investment and a commodity terms of trade index. Appendix \ref{sec:appendix_aggregate_data} describes the sources of our variables and further details of our datasets. 

Lastly, we describe how we estimate and quantify the impact of a US monetary policy tightening shock on Emerging Market economies through the pooled panel SVAR model. We order the vectors in $y_t$ such that the variables in $m_t$ are ordered first and have contemporaneous impact on the rest of the variables in the model. Consequently, this identification scheme takes the exogenous shocks as given and focuses on estimating the impact on Emerging Market variables. In particular, we use the US monetary policy shocks constructed by \cite{bauer2022reassessment}. These shocks are constructed by isolating the component of high-frequency interest rates around FOMC meetings which are orthogonal to macroeconomic and financial data available at the moment.\footnote{In \cite{bauer2022reassessment}, argue that using high-frequency financial surprises around FOMC meetings may be contaminated with other shocks which occur simultaneously. Furthermore, the authors expand the set of monetary policy announcements to include speeches by the Fed Chair, which essentially doubles the number and importance of announcements in their dataset.}$^{,}$\footnote{In Appendix \ref{sec:appendix_model_svar_results} we show that results are robust to identifying US tightening shocks using only the high-frequency surprise of the Federal Funds Futures around FOMC meetings. As in \cite{bauer2022reassessment}, results are substantially larger and more significant when using the shocks which clean any potential contamination.} This shock is ordered first in our set of variables and the structural shocks are recovered through a standard Choleski identification scheme.

The models are estimated using Bayesian tools with a standard Normal-Wishart prior over the auto-regressive coefficient and the innovation volatilities and using two lags, i.e., $p=2$. The hyper-parameter which govern the overall tightness and lag decay of the variance prior are set to 0.1 and 1 respectively, typical values on the literature (see \cite{dieppe2016bear}). The model results are based on 12,000 iterations of the Gibbs sampler, discarding the first 2,000 simulations for convergence.\footnote{Details of the model's estimation procedure, prior and posterior computation are in Appendix \ref{sec:appendix_model_details}.} 

\noindent
\textbf{Panel SVAR results.} We start by studying the impact of a US tightening of interest rates on Emerging Market economies. To do so, we estimate the impulse response functions of a US tightening shock through a panel SVAR model. 

\begin{figure}[ht]
    \centering
    \caption{IRF Analysis - US Tightening Shock \\ \footnotesize Panel SVAR -  Response of Domestic Variables}
    \label{fig:Only_Domestic_Panel_Larger_Norm}    
    \includegraphics[width=14cm,height=12cm]{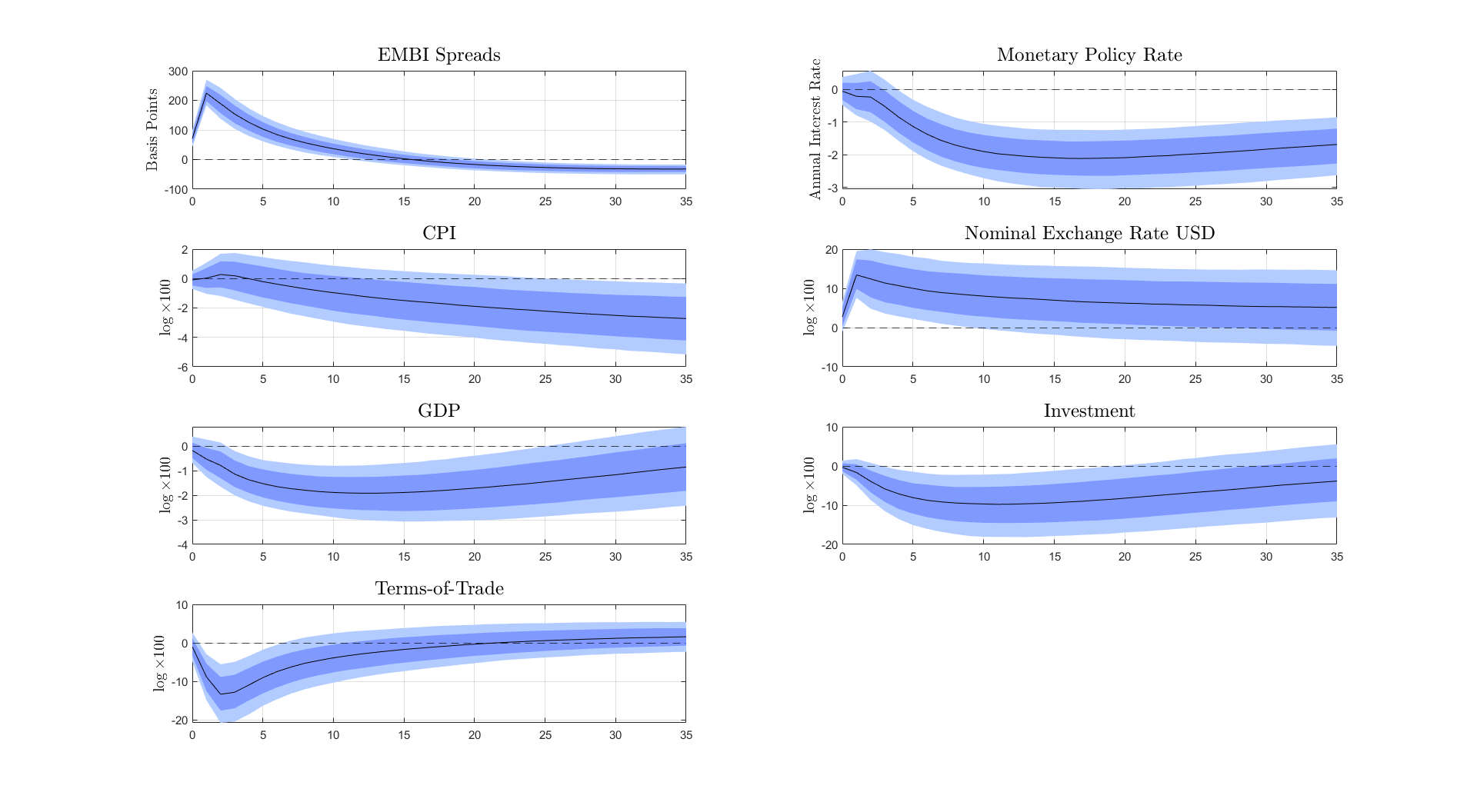}
    \floatfoot{\textbf{Note}: Median (black line), (68\% confidence interval in dark blue band), (90\% confidence interval in light blue band). Panels' order starts from the top left corner, moving from left to right, and later from top to bottom. The impulse response functions are computed from a exogenous US tightening shock which causes the 10-Year Treasury yield increase by 50 basis points.}
\end{figure}

Figure \ref{fig:Only_Domestic_Panel_Larger_Norm} presents the impact of a US tightening shock that leads to a 50 basis point increase in the 10 year US treasury yield on the panel of Emerging Market economies. Overall, results are in line with a traditional literature which study the spillovers of US interest rate shocks (see \cite{vicondoa2019monetary}) and with a recent literature which measure the impact of pure US monetary policy shocks (see \cite{camara2021spillovers}).
Panel 1.1 shows that EMBI spreads show a sharp increase, peaking on the period following the shock above 200 basis points above pre-shock levels. The domestic monetary policy does not exhibit significative response in the first quarter of the shock and later shows a drop of between 1\% and 2\%. Observing panels 2.1 and 2.2 together, the shock leads to a real exchange rate depreciation slightly above 10\%. The third row shows the impact of a US tightening shock on real activity. Panel 3.1 shows that GDP drops in a hump-shaped way reaching a 2\% drop around 10 months after the initial shock. Panel 3.2 shows that the drop in GDP is accompanied by a sharp drop  in aggregate investment which reaches a drop of 10\% to pre-shock levels around 10 months after the initial shock. Given that the share of aggregate investment in GDP in these countries is close to 20\%, a quick back-of-the-envelope calculation suggests that the drop in investment is a main driver of the drop in GDP. Finally, Panel 4.1 shows that a US tightening shock also triggers a drop in countries' terms-of-trade which may exacerbate the economic downturn.  

\noindent
\textbf{Results using expenditure shares.} To provide further evidence of the role of investment explaining the drop in GDP in response to a US interest rate tightening, we re-estimate the panel SVAR model using GDP's expenditure shares. In particular, we keep the vector of $y^{*}_t$ unchanged and re-define the vector of variables $y_t$ as: (i) EMBI spreads, (ii) Monetary Policy Rate, (iii) CPI, (iv) nominal exchange rate, (v) GDP, (vi) Terms-of-Trade, (vii) net exports to GDP share, (viii) consumption to GDP share, (ix) government to GDP  share, (x) investment to GDP share. The latter four variables are constructed by computing the shares at the quarterly frequency and interpolating to the monthly frequency. 

\begin{figure}[ht]
    \centering
    \includegraphics[width=12cm,height=9cm]{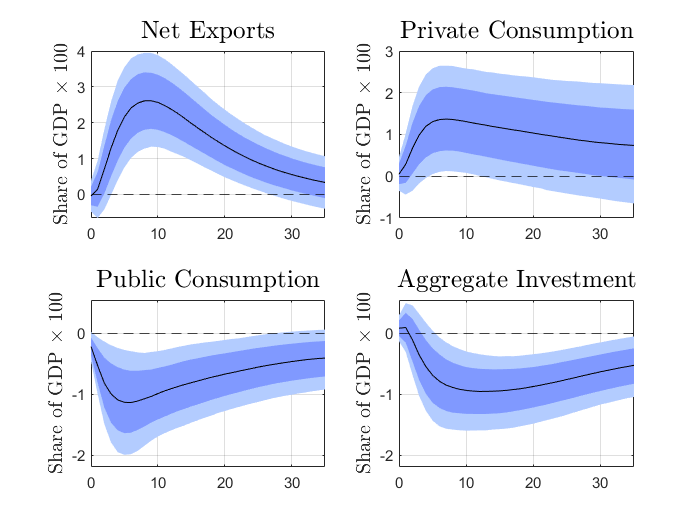}
    \caption{IRF Analysis - Expenditure Shares of GDP}
    \label{fig:Full_Panel_Shares}
    \floatfoot{\textbf{Note}: Median (black line), (68\% confidence interval in dark blue band), (90\% confidence interval in light blue band). Panels' order starts from the top left corner, moving from left to right, and later from top to bottom. The impulse response functions are computed from a exogenous US tightening shock which causes the 10-Year Treasury yield increase by 50 basis points.}
\end{figure}

Figure \ref{fig:Full_Panel_Shares} presents the dynamics of the four shares of expenditure to total GDP. Panel 1.1 shows that a US monetary policy shocks leads to a sharp increase in the share of net exports over total GDP. This improvement of the trade balance is in line with tighter international financial conditions and capital outflows. Panel 1.2 shows that the share of private consumption over total GDP increases in response to a US monetary policy shock. 
At a glance, this result may seem at odds with conventional wisdom. While higher foreign interest rates may trigger a drop in consumption due to intertemporal substitution, this alone does not imply that consumption should fall by more than output. Furthermore, this result suggests that consumption falls by relatively less than total GDP, as households seek to smooth their consumption profile. Panels 2.1 and 2.2 show the two components of aggregate expenditure explaining the drop in aggregate GDP. These panels show that the share of public consumption and aggregate investment drop by 1 percentage point each. The drop in public consumption can be explained by the increase in EMBI spreads which surely curtails the sovereign's ability to finance government expenditure.\footnote{An additional potential explanation to the drop in public consumption is that the vast majority of public consumption is non-tradable goods. Given the sharp nominal and real depreciation of the exchange rate, the relative price of these goods will sharply decline.} The drop in the share of investment is persistent, with tight confidence intervals different than zero even 3 years after the initial shock.

\begin{figure}[ht]
    \centering
    \caption{IRF Analysis - US Tightening Shock \\ \footnotesize Panel SVAR -  High vs. Low EMBI Countries}
    \label{fig:High_Low_EMBI_Comparison_GDP}    
    \includegraphics[scale=0.45]{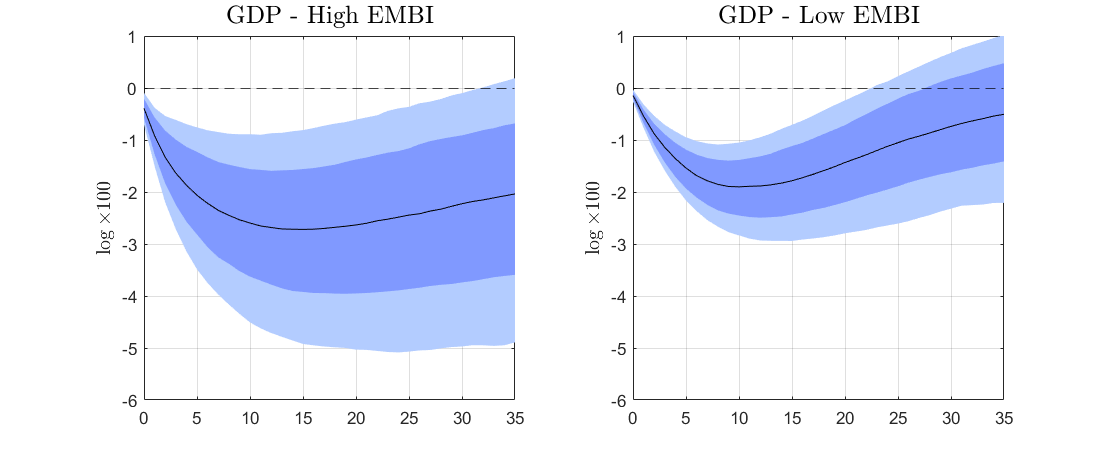}
    \floatfoot{\textbf{Note}: Median (black line), (68\% confidence interval in dark blue band), (90\% confidence interval in light blue band). The impulse response functions are computed from a exogenous US tightening shock which causes the 10-Year Treasury yield increase by 50 basis points. The left panel presents the results for the ``High EMBI Countries'', the right panel presents the results for the ``Low EMBI Countries''.}
\end{figure}

\noindent
\textbf{Role of financial conditions across countries.} 
Although we study in detail the role of tighter financial conditions in Section \ref{sec:firm_evidence} by exploiting heterogeneity in firms' financial position, we test whether there is aggregate-evidence in support of our hypothesis. To do so, we divide our sample of seven countries between ``Low EMBI Countries'' (Chile, Peru and Mexico) and ``High EMBI Countries'' (Brazil, Indonesia, Turkey and South Africa). These division is computed by computing the average monthly EMBI spreads for each country for the sample period of the panel SVAR model and ordering the countries. We estimate the model for each sample separately under our benchmark specification. Figure \ref{fig:High_Low_EMBI_Comparison_GDP} shows the response of real GDP for both samples.
The median response, plotted in black solid line in both sub-plots, shows a greater drop in GDP for countries in the ``High EMBI'' sample than for ``Low EMBI'' sample. In particular, the drop in GDP for the first sample peaks 12 months after the shock, close to 3\% below pre-shock levels. For the latter sample, the drop in GDP peaks between 8 and 10 months after the shock, close to 2\% below pre-shock levels. While this result is not enough to argue that tighter financial conditions casually lead to a greater impact of US interest rate shocks it serves as aggregate motivation for our firm-level analysis in Section \ref{sec:firm_evidence}.\footnote{Similar exercise of estimating the impact of US tightening shocks on sub-samples of countries also find that countries which exhibit tighter financial conditions experience a greater impact in response to US interest rate shocks. See \cite{dedola2017if} for instance.}

In summary, the results presented above suggest that a US interest rate hike causes an economic recession driven by lower aggregate investment. In Appendix \ref{sec:appendix_model_svar_results}, we presents for the full set of variables in the SVAR, i.e. the dynamics of US variables, and under different variable-specifications. In Appendix \ref{sec:appendix_model_svar_results} we also show that these results hold under different robustness checks, including substituting the interpolated real GDP with a monthly produced industrial production index and other identification schemes. 

While the panel SVAR model allows us to causally identify the negative impact of a US tightening shock on Emerging Market economies, it does not allow us to identify the relevance of the different possible transmission channels. For instance, a US interest rate hike might lead to tighter financial conditions in Emerging Markets. On the one hand, the increase in EMBI spreads suggest tighter financial conditions. On the other hand, the lack of a domestic monetary policy rate hike may suggest the opposite. Thus, it is not straightforward that tighter financial conditions are the main driver of the economic recession. Similarly, an economic recession accompanied with a drop in investment can be explained by a transitory negative wealth effect triggered by a drop in terms-of-trade (see Panel 4.1).

%%%%%%%%%%%%%%%%%%%%%%%%%%%%%%%%%%%%%%%%%%%%%%%%%%%%%%%%%%%%%
\subsection{Evidence for Chile} \label{subsec:aggregate_evidence_chile}

Next, we estimate a standard SVAR model for Chile to test whether the previous results for the panel of Emerging Markets are also present for this country. This will serve as aggregate motivation for our firm-level analysis of the transmission of US interest rate shocks in Section \ref{sec:firm_evidence}.

The model estimated is a standard Bayesian SVAR model. The set of variables is identical to that used for the panel SVAR model. We use a Minnesota prior on the model's parameters.\footnote{We choose hyper-parameters values of 0.80 for the auto-regressive coefficients, 0.10 for overall tightness, 0.50 for the cross-variable weighting and lag decay of 1. The model is estimated using 6 lags.} Once again, identification of US interest rate shocks is carried out using standard Choleski factorization and ordering the \cite{bauer2022reassessment} shocks first in our set of variables. 

\begin{figure}[ht]
    \centering
    \caption{IRF Analysis - US Tightening Shock \\ \footnotesize Chile -  Response of Domestic Variables}
    \label{fig:Chile_OnlyDomestic_Larger}    
    \includegraphics[width=14cm,height=12cm]{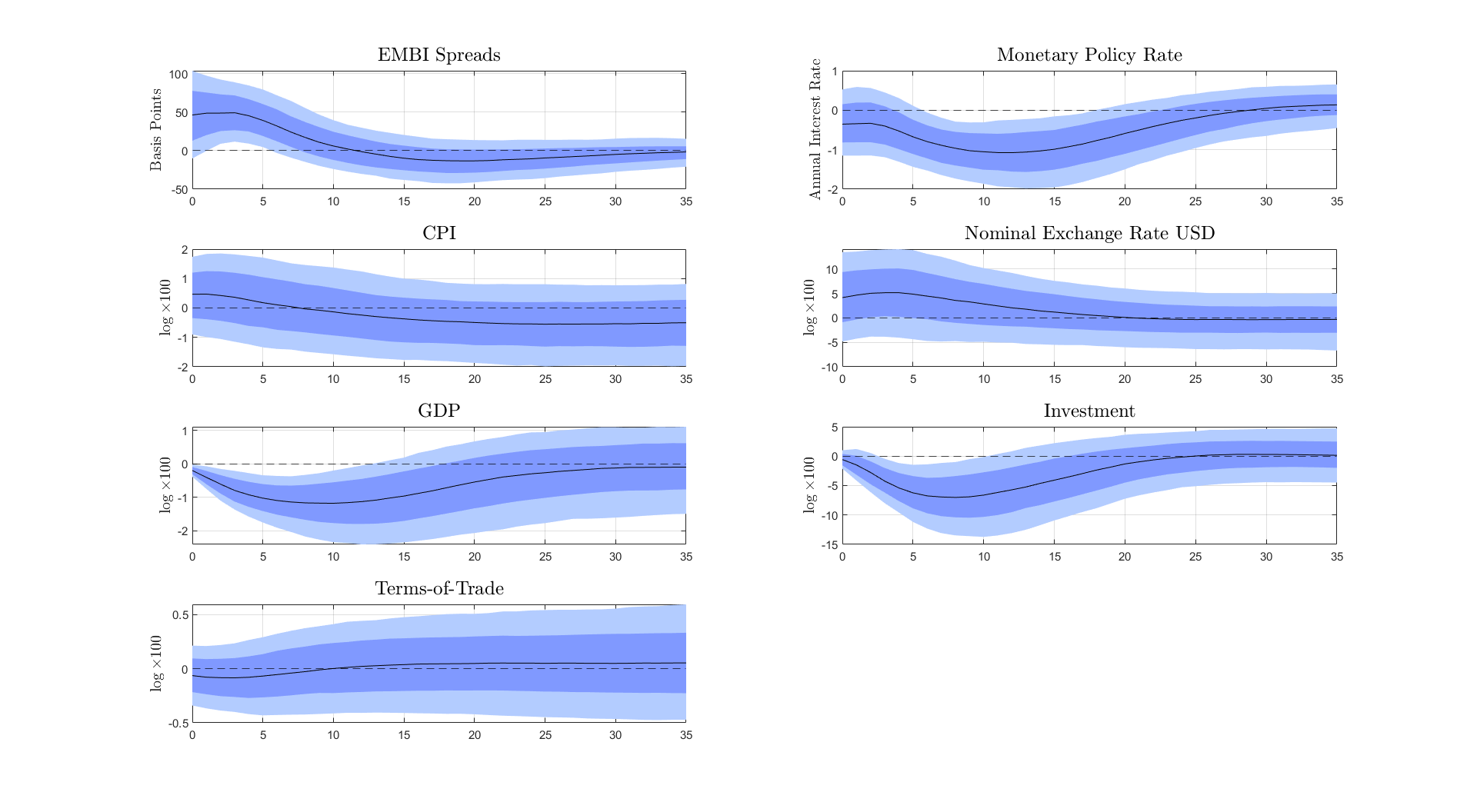}
    \floatfoot{\textbf{Note}: Median (black line), (68\% confidence interval in dark blue band), (90\% confidence interval in light blue band). Panels' order starts from the top left corner, moving from left to right, and later from top to bottom. The impulse response functions are computed from a exogenous US tightening shock which causes the 10-Year Treasury yield increase by 50 basis points.}
\end{figure}

Figure \ref{fig:Chile_OnlyDomestic_Larger} presents the impact of a US tightening shock presents that leads to a 50 basis point increase in the 10 year US treasury yield on the Chilean economy. Altogether, while less-accurate (wider uncertainty binds) results are in line with the panel-SVAR results presented in Figure \ref{fig:Only_Domestic_Panel_Larger_Norm}. 
Panel 1.1 shows that EMBI spreads show a sharp increase on impacts close to 50 basis points above pre-shock levels. Panel 1.2 shows that the domestic monetary policy does not react significantly to a US tightening shock on impact, and shows a slight reduction between 6 and 12 months after the initial shock. Again, Panels 2.1 and 2.2 show that a US tightening shock leads to a real exchange rate depreciation close to 5\%. Panels 3.1 and 3.2 show that as it was the case for the panel of countries, a US tightening shock leads to a significative reduction of GDP driven by a sharp drop in aggregate investment. 

\begin{figure}[ht]
    \centering
    \caption{IRF Analysis - US Tightening Shock \\ \footnotesize Chile -  Response of Investment Components}
    \label{fig:Investment_Components_Large}
    \includegraphics[width=15cm,height=10cm]{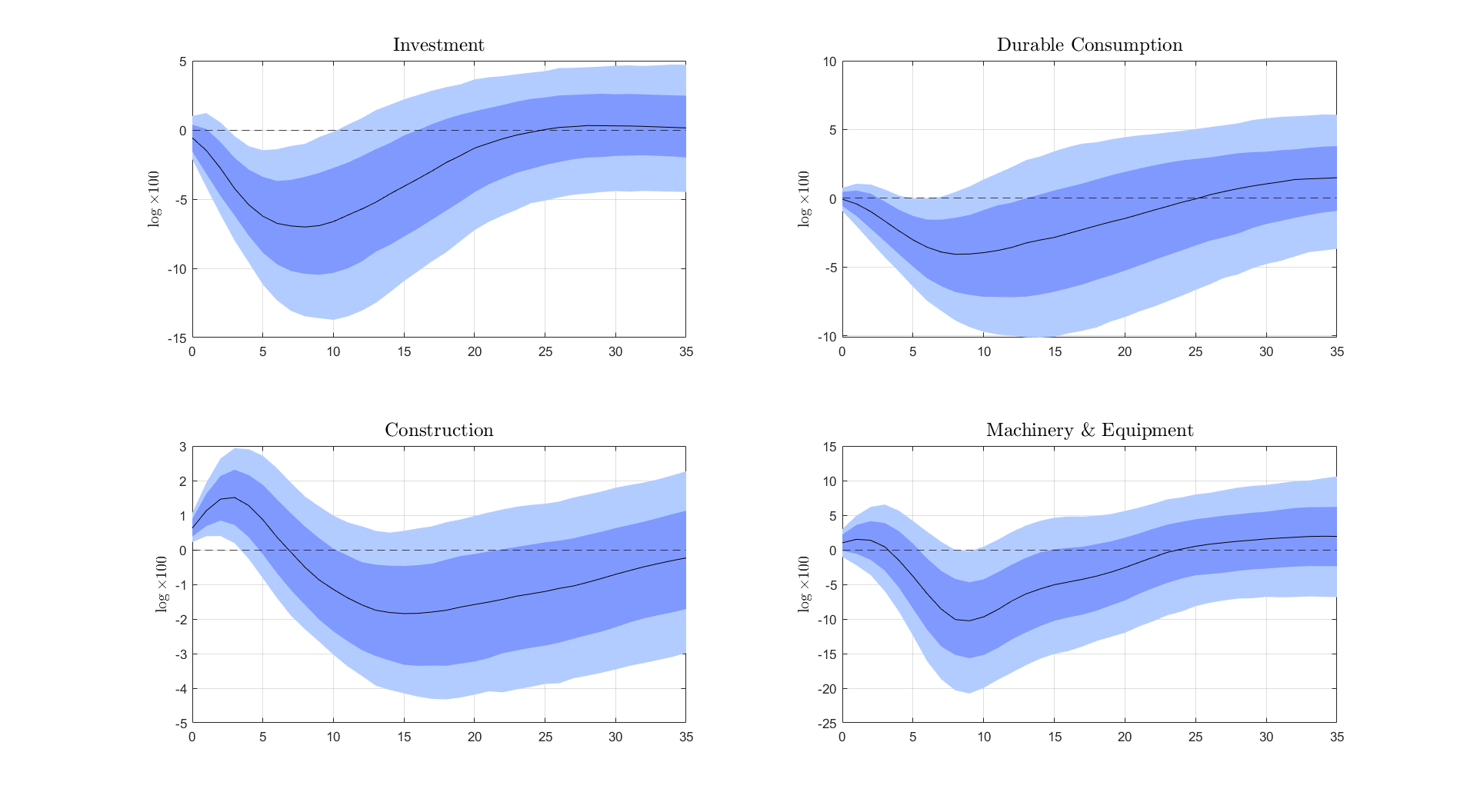}
    \floatfoot{\textbf{Note}: Median (black line), (68\% confidence interval in dark blue band), (90\% confidence interval in light blue band). Panels' order starts from the top left corner, moving from left to right, and later from top to bottom. The impulse response functions are computed from a exogenous US tightening shock which causes the 10-Year Treasury yield increase by 50 basis points. Each panel shows the dynamics of a different estimated model. Panel 1.1 shows the response of aggregate investment under the benchmark specification. Panel 1.2, 2.1 and 2.2 show the response of durable consumption, a construction index and a firms' machinery and equipment index by substituting out aggregate investment in the benchmark specification.}
\end{figure}

Lastly, we test which components of aggregate investment are the most responsive to a US tightening shock. We test this by estimating our benchmark model specification but replacing aggregate investment with three alternative variables: (i) a durable consumption index, (ii) a construction index, (iii) a firm's machinery and equipment purchase index. Figure \ref{fig:Investment_Components_Large} presents the dynamics of aggregate investment (already shown in Figure \ref{fig:Chile_OnlyDomestic_Larger}) and the three components to a US tightening shock.   
The overall message of this figure is that the fall in aggregate investment is primarily explained by a sharp reduction in firms' purchases of machinery and equipment. Panel 1.2 shows that durable consumption falls in response to a US tightening shock, reaching a gap of 4\% to pre-shock levels 8 months after the initial shock. Panel 2.1 shows that the construction index actually increases during the first 5 months after the initial shock, to later show a drop with wide uncertainty bounds. Lastly, Panel 2.2 shows that a US interest rate shock that firms' purchase of machinery and equipment experiences a sharp and significant drop, reaching a fall (at the median) of 10\% below pre-shock levels 8 months after the initial shock.

In summary, this section presented aggregate evidence that a US tightening shock leads to a persistent drop in economic output. We do so by estimating both a panel and a Chile specific SVAR model and computing impulse response functions. We also argue that the sharp drop in aggregate investment is a main driver of the economic recession. Motivated by these aggregate facts, in the next section we test the importance of the different transmission channels of a US tightening shock, particularly that of financial heterogeneity, by using firm-level data.

%%%%%%%%%%%%%%%%%%%%%%%%%%%%%%%%%%%%%%%%%%%%%%%%%%%%%%%%%%%%%%%%%%%%%%
%% Firm Level Evidence
\section{Firm Level Evidence} \label{sec:firm_evidence}

In this section we introduce the main empirical results of the paper. First, we describe the micro level dataset and present key stylized facts on firms' capital accumulation and balance sheet dynamics. Second, we use the high-frequency movements in the Fed Funds Futures constructed in Section \ref{sec:aggregate_evidence} to quantify the impact of a US monetary tightening on firm level capital accumulation. Additionally, we show that results are heterogeneous depending on a firms' indebtedness.

\noindent
\textbf{Data description.} We construct our sample of firm level variables from the quarterly CMF (\textit{Comision para el Mercado Financiero}) portal, a panel of publicly listed Chilean firms. This dataset satisfies several key requirements for our study: it is at a quarterly frequency, an enough high frequency to study the impact of monetary policy and shocks coming from the \textit{Rest of the World}; it is a long panel covering multiple decades, allowing us to use within-firm variation in our empirical specifications; contains rich balance-sheet information, such as assets, liabilities (in both domestic and foreign currency), allowing us to construct key variables of interest. To the best of our knowledge, the use of this dataset is novel in the analysis of US monetary policy shocks in emerging open economies. Table \ref{tab:firms_by_sectors} in Appendix \ref{sec:appendix_data_firms} presents the distribution of firms by their self-reported sector of activities.

Table \ref{tab:firms_summary_statistics} presents summary statistics about firms' sales growth, capital accumulation, leverage and currency-mismatch. 
\begin{table}[ht]
    \centering
    \caption{Firms' Summary Statistics}
    \label{tab:firms_summary_statistics}
    \footnotesize
    \begin{tabular}{l c c c c c c c}
	&	Mean	&	p10	&	p25 	&	Median	&	p75 	&	p90	&	SD	\\ \hline \hline
%Sales Growth $\log py_{i,t}$	&	0.039	&	-1.389	&	-0.255	&	0.358	&	0.532	&	0.743	&	0.898	\\
%Sales Growth $\log py_{i,t+4}$	&	0.128	&	-1.332	&	-0.270	&	0.434	&	0.690	&	0.933	&	1.048	\\
%\\
Investment Growth $\log \Delta k_{i,t} $	&	0.029	&	-0.051	&	-0.010	&	0.017	&	0.054	&	0.115	&	0.196	\\
Investment Growth $\log \Delta k_{i,t+4} $	&	0.136	&	-0.093	&	0.004	&	0.090	&	0.221	&	0.412	&	0.406	\\
\\
Leverage: Liab. - Assets	&	0.481	&	0.077	&	0.242	&	0.434	&	0.594	&	0.770	&	0.862	\\
Leverage: Short-Term Liab. - Assets	&	0.219	&	0.027	&	0.070	&	0.157	&	0.279	&	0.454	&	0.313	\\
\\
Currency-Mismatch & -0.025	&	-0.311	&	-0.020	&	0.000	&	0.015	&	0.181	&	0.242 \\ \hline \hline
    \end{tabular}
    \floatfoot{\textbf{Note:} The statistics computed above are constructed for our whole sample. In the first and second row sales are approximated as firms' ``Revenue from ordinary activities''. In the last row ``Currency-Mismatch'' is computed as $(\text{Foreign Currency Liabilities} - \text{Foreign Currency Assets})/(\text{Total Assets})$.}
\end{table}
Our two main measures of investment are $\Delta \log k_{i,t}=\log k_{i,t}-\log k_{i,t-1}$ and $\Delta \log k_{i,t+4}=\log k_{i,t+4}-\log k_{i,t-1}$ which represent firm $i$'s log growth rate of the book value of capital from in period $t$ and $t+4$ compared to period $t-1$, respectively. We use two measures of firms' leverage as proxies of indebtedness. First, we measure leverage as the ratio of a firm's total liabilities to total book value of assets. Total liabilities are comprised of both short and long term liabilities and both domestic and foreign currency liabilities. Second, we measure leverage as the ratio of a firm's short-term liabilities to total book value of capital. We focus on short-term liabilities (under one year) as they represent, on average, more than 50\% a firms' total liabilities.\footnote{For our total sample of firms, the average ratio of short-term liabilities to total liabilities is equal to  $52.59\%$, with percentiles 25 and 75 equal to 26.73\% and 78.99\%, respectively.} Finally, we construct a measure of firms' exposure to currency-mismatch, defined as the the ratio of the difference of between foreign currency liabilities and foreign currency assets to total assets.

\noindent
\textbf{Firm level results.} We turn to estimating the impact of a US interest rate shock on firms' investment. Our baseline empirical specification is 
\begin{align} \label{eq:baseline_heterogeneous}
    \log k_{i,t} - \log k_{i,t-1} = \Delta \log k_{i,t} = \alpha_i + \alpha_{s,t} + \beta \left(l_{i,t-1} - \mathbb{E}_{i} \left[ l_{t}\right] \right) \epsilon^{X}_t + \Gamma' Z_{i,t-1}  + e_{i,t} 
\end{align}
where $\alpha_i$ is a firm fixed effect, $\alpha_{s,t}$ is a sector-time fixed effect, $\epsilon^{X}_t$ is the simple quarterly average of the shock constructed used in Section \ref{sec:aggregate_evidence}, $\left(l_{i,t-1} - \mathbb{E}_{i} \left[ l_{t}\right] \right)$ captures within firm $i$ variation in leverage, $Z_{i,t-1}$ is a vector of controls, and $e_{i,t}$ is a residual. The main coefficient of interest is $\beta$, which measures how the semi-elasticity of investment $\Delta \log k_{i,t}$ with respect to US monetary policy shocks depends on a firm's financial position.

The empirical specification in Equation \ref{eq:baseline_heterogeneous} controls for a number of factors which may simultaneously affect firms' investment and financial position. The firm fixed effect $\alpha_i$ capture permanent differences in investment behavior across firms. The sector-time fixed effect $\alpha_{s,t}$ capture differences in how broad sectors are exposed to aggregate shocks. The vector $Z_{i,t-1}$ includes the firms' past financial position $l_{i,t-1}$, lagged asset growth, and lagged currency mismatch.

Table \ref{tab:baseline_heterogeneous} presents the results of estimating our baseline specification in Equation \ref{eq:baseline_heterogeneous}. We carry out a normalization to simplify the interpretation of parameter $\beta$. We standardized the firm's demeaned leverage $\left(l_{i,t-1} - \mathbb{E}_{i} \left[ l_{t}\right] \right)$ over the entire sample, i.e., using firm specific standard deviations. This implies that the units of $\beta$ are expressed in terms of standard deviations. 
\begin{table}[ht]
    \centering
    \caption{Heterogeneous Response of Investment  to $\epsilon^{X}_t$}
    \footnotesize
    \label{tab:baseline_heterogeneous}
\begin{tabular}{lcccc}
 & \multicolumn{4}{c}{Firm investment - $\Delta \log k_{i,t}$ } \\
  & (1) & (2) & (3) & (4) \\ \hline \hline
 &  &  &  &  \\
$\left(l_{i,t-1} - \mathbb{E}_{i} \left[ l_{t}\right] \right) \epsilon^{X}_t$ & -0.380*** & -0.387*** & -0.296** & -0.306** \\
 & (0.127) & (0.129) & (0.141) & (0.142) \\
 &  &  &  &  \\
Firm FE & yes & yes & yes & yes \\
Sector-Time FE & yes & yes & yes & yes \\
Control for Lagged Inv. & no & yes & yes & yes \\
Control for Lagged Currency Mismatch & no & no & yes & yes \\
Control for $l_{i,t-1}$ & no & no & no & yes \\
 &  &  &  &  \\
Observations & 12,646 & 12,473 & 11,967 & 11,869 \\  \hline
 $R^{2}$ &  0.143 & 0.153 & 0.150 & 0.163 \\ \hline \hline
\multicolumn{5}{c}{ Clustered standard errors in parentheses} \\
\multicolumn{5}{c}{ *** p$<$0.01, ** p$<$0.05, * p$<$0.1} \\
\end{tabular}
\floatfoot{\textbf{Note:} Results from estimating $\log k_{i,t} - \log k_{i,t-1} = \alpha_i + \alpha_{s,t} + \beta \left(l_{i,t-1} - \mathbb{E}_{i} \left[ l_{t}\right] \right) \epsilon^{X}_t + \Gamma' Z_{i,t-1}  + e_{i,t}$. We have standardized $\left(l_{i,t-1} - \mathbb{E}_{i} \left[ l_{t}\right] \right)$ over the entire sample using firm specific standard deviations. Standard errors are clustered at the firm and time level.  }
\end{table}
Across all columns of Table \ref{tab:baseline_heterogeneous}, firms which are more levered in period $t-1$ experience a greater drop in investment after a US interest rate shock than relatively less levered firms. Column (1) implies that a firm has a 0.2 impact of a US interest rate shock when it is a one standard deviation more levered than it typically is in our sample. Columns (2)-(4) show that introducing firm-controls only makes the estimated impact on highly-levered firms even greater. Column (4) shows that a one standard deviation increase in firms leverage leads to 0.306 greater impact of US monetary policy shocks on their investment.

Next, we introduce an alternative empirical specification which estimates how financial heterogeneity matters for the transmission of US monetary policy shocks on firms' investment. To do so, we estimate the following empirical specification
\begin{align} \label{eq:baseline_heterogeneous_average}
    \Delta \log k_{i,t} = \alpha_i + \alpha_{s,q} + \gamma \epsilon^{X}_t + \beta \left(l_{i,t-1} - \mathbb{E}_{i} \left[ l_{t}\right] \right) \epsilon^{X}_t + \Gamma'_1 Z_{i,t-1} + \Gamma'_2 Y_{t-1}  + \tilde{e}_{i,t} 
\end{align}
where we remove the sector-time fixed effects, $\alpha_{s,q}$ is a sector $s$, quarter $q$ seasonal fixed effect, and $Y_{t-1}$ is a vector with four lags of the annualized inflation rate, the log of GDP and the log of the nominal exchange rate with the US dollar. 

Table \ref{tab:baseline_heterogeneous_average} presents the results of this alternative specification. Column (1) shows that the average investment semi-elasticity if close to 0.12 and not significantly different from zero. However, a firm with a leverage one standard deviation that the sample average exhibits a semi-elasticity equal to -0.371.
\begin{table}[ht]
    \centering
    \caption{Heterogeneous Response of Investment to $\epsilon^{X}_t$ \\ \footnotesize Alternative Specification}
    \label{tab:baseline_heterogeneous_average}
    \footnotesize
\begin{tabular}{lcccc}
 & \multicolumn{4}{c}{Firm investment - $\Delta \log k_{i,t}$ } \\
  & (1) & (2) & (3) & (4) \\ \hline \hline
 &  &  &  &  \\
$\epsilon^{X}_t$  & 0.120 & 0.128 & 0.126 & 0.104 \\
 & (0.143) & (0.146) & (0.144) & (0.119) \\ 
 \\
$\left(l_{i,t-1} - \mathbb{E}_{i} \left[ l_{t}\right] \right) \epsilon^{X}_t$ & -0.371*** & -0.375*** & -0.294** & -0.275** \\
 & (0.116) & (0.120) & (0.135) & (0.134) \\
 &  &  &  &  \\
Firm FE & yes & yes & yes & yes \\
Sector-Time FE & yes & yes & yes & yes \\
Control for Lagged Inv. & no & yes & yes & yes \\
Firm-Controls & no & no & yes & yes \\
Aggregate-Controls & no & no & no & yes \\
 &  &  &  &  \\
Observations & 12,646 & 12,473 & 11,869 & 11,863 \\ \hline
$R^{2}$ & 0.057 & 0.065 & 0.076 & 0.081 \\ \hline \hline
\multicolumn{5}{c}{ Clustered standard errors in parentheses} \\
\multicolumn{5}{c}{ *** p$<$0.01, ** p$<$0.05, * p$<$0.1} \\
\end{tabular}
\floatfoot{\textbf{Note:} Results from estimating $\Delta \log k_{i,t} = \alpha_i + \alpha_{s,q} + \gamma \epsilon^{X}_t + \beta \left(l_{i,t-1} - \mathbb{E}_{i} \left[ l_{t}\right] \right) \epsilon^{X}_t + \Gamma'_1 Z_{i,t-1} + \Gamma'_2 Y_{t-1}  + \tilde{e}_{i,t}$. We have standardized $\left(l_{i,t-1} - \mathbb{E}_{i} \left[ l_{t}\right] \right)$ over the entire sample using firm specific standard deviations. Standard errors are clustered at the firm and time level.  }
\end{table}
Columns (2)-(4) shows that introducing firm and aggregate controls has no impact on the average semi-elasticity $\gamma$ and but maintains the main results over the differential impact of firms with relatively higher leverage. Column (4) shows that with firm and aggregate controls, the differential impact is close to -0.30. In consequence, our interaction coefficients imply an economically and quantitative meaningful degree of heterogeneity. 

Lastly, we test whether the heterogeneous impact of US interest rate shocks on firms' investment persists across time. To estimate the dynamics of these differential responses across firms, we estimate dynamic versions of the specifications in Equation \ref{eq:baseline_heterogeneous} and \ref{eq:baseline_heterogeneous_average} motivated by \cite{jorda2005estimation} local projection method. In particular, we estimate
\begin{align} \label{eq:dynamics_first}
    \Delta \log k_{i,t+j} &=  \alpha_{i,j} + \alpha_{s,t,j} + \beta_j \left(l_{i,t-1} - \mathbb{E}_{i} \left[ l_{t}\right] \right) \epsilon^{X}_t + \Gamma_{h} Z_{i,t-1}  + e_{i,t+j}
\end{align}
\begin{align} \label{eq:dynamics_second}
    \Delta \log k_{i,t+j} &=  \alpha_{i,j} + \alpha_{s,q,j} + \gamma \epsilon^{X}_t + \beta_j \left(l_{i,t-1} - \mathbb{E}_{i} \left[ l_{t}\right] \right) \epsilon^{X}_t + \Gamma'_{1,h} Z_{i,t-1} + \Gamma'_{2,h} Y_{i,t-1} + e_{i,t+j}
\end{align}
where $\log k_{i,t+j} - \log k_{i,t-1}$ and $j \geq 1$ indexes the forecast horizon. Under this specification, the coefficient $\beta_j$ measures how the cumulative response of investment in period $t+j$ to a US interest rate shock in period $t$ depends on a firms' standardized leverage in period $t-1$. We use the cumulative change in capital on the left-hand side of Equations \ref{eq:dynamics_first} and \ref{eq:dynamics_second} in order to easily assess the implications of our estimates for the capital stock itself.

\begin{figure}[ht]
     \centering
     \caption{Dynamics of Differential Responses to $\epsilon^{X}_t$}
     \label{fig:baseline_dynamics}
     \begin{subfigure}[b]{0.475\textwidth}
         \centering
         \includegraphics[width=\textwidth]{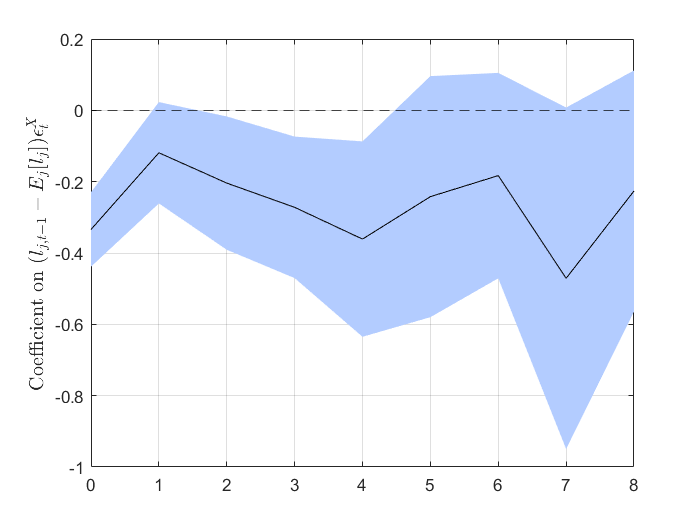}
         \caption{First Specification \footnotesize Eq. \ref{eq:dynamics_first}}
         \label{fig:Interaction_First}
     \end{subfigure}
     \begin{subfigure}[b]{0.475\textwidth}
         \centering
         \includegraphics[width=\textwidth]{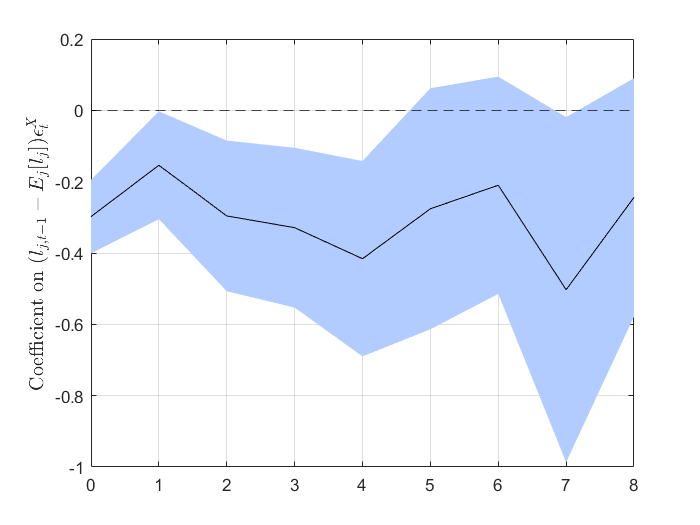}
         \caption{Second Specification \footnotesize Eq. \ref{eq:dynamics_second}}
         \label{fig:Interaction_Second}
     \end{subfigure}
     \floatfoot{\textbf{Note:} The left panel presents results of estimating equation $\Delta \log k_{i,t+j} =  \alpha_{i,j} + \alpha_{s,t,j} + \beta_j \left(l_{i,t-1} - \mathbb{E}_{i} \left[ l_{t}\right] \right) \epsilon^{X}_t + \Gamma_{h} Z_{i,t-1}  + e_{i,t+j}$. The right panel presents results of estimating equation $\Delta \log k_{i,t+j} =  \alpha_{i,j} + \alpha_{s,q,j} + \gamma \epsilon^{X}_t + \beta_j \left(l_{i,t-1} - \mathbb{E}_{i} \left[ l_{t}\right] \right) \epsilon^{X}_t + \Gamma'_{1,h} Z_{i,t-1} + \Gamma'_{2,h} Y_{i,t-1} + e_{i,t+j}$. The black line presents the point estimate of parameter $\beta_j$ for $j = \{0,\ldots,8\}$. The light blue shade presents 90\% confidence intervals. The standard errors are clustered at the firm level. }
\end{figure}

Figure \ref{fig:baseline_dynamics} presents the estimated coefficients for $\beta_j$ for both specifications, Equations \ref{eq:dynamics_first} and \ref{eq:dynamics_second}, for $j=\{0,\ldots,8\}$. To begin with, this figure shows that firms with relatively high leverage are consistently more responsive to a US interest rate shock. The estimated coefficients are persistently significantly different from zero up to four quarters after the initial shock, with the uncertainty bounds becoming increasingly larger from $j=5$ onward. Furthermore, the point estimates are similar in magnitudes across panels and/or specifications, implying a semi-elasticity close to -0.4 for a firm with a leverage one standard deviation above its sample mean. 

\noindent
\textbf{Additional Empirical Results.} In Appendix \ref{sec:appendix_additional_firm_level_results} we carry out several additional empirical exercises. First, we estimate whether results are robust to different specifications of the interaction term between the US interest rate shock and firms' relative indebtedness. To do so, we define two alternative specifications which aim to capture firms' indebtedness. In particular, we replace the interaction term
\begin{align*}
    \left(l_{i,t-1} - \mathbb{E}_{i} \left[ l_{t}\right] \right) \epsilon^{X}_t
\end{align*}
with two interaction terms constructed using indicator functions
\begin{align*}
    \mathbbm{1} \left[\frac{l_{i,t} - \mathbb{E}_i \left[l_t \right]}{\sigma_{l} \left( l_t\right)}  > 0 \right] & \times \epsilon^{X}_t \\
    \mathbbm{1} \left[\frac{l_{i,t} - \mathbb{E}_i \left[l_t \right]}{\sigma_{l} \left( l_t\right)}  > 1 \right] & \times \epsilon^{X}_t
\end{align*}
where the first specification is an indicator function that takes the value of 1 if the standardized present measure of firm leverage is above its sample mean and zero otherwise, and the second specification is an indicator function which takes the value of 1 if the standardized present measure of firm leverage is one standard deviation above its sample mean and zero otherwise. We estimate the following empirical specifications
\begin{align} 
     \Delta \log k_{i,t} &= \alpha_i + \alpha_{s,t} + \beta \mathbbm{1} \left[\frac{l_{i,t} - \mathbb{E}_i \left[l_t \right]}{\sigma_{l} \left( l_t\right)}  > \bar{c} \right] \times \epsilon^{X}_t + \Gamma' Z_{i,t-1}  + e_{i,t} \label{eq:indicator_heterogeneous} \\
    \Delta \log k_{i,t} &= \alpha_i + \alpha_{s,q} + \gamma \epsilon^{X}_t + \beta \mathbbm{1} \left[\frac{l_{i,t} - \mathbb{E}_i \left[l_t \right]}{\sigma_{l} \left( l_t\right)}  > \bar{c} \right] \epsilon^{X}_t + \Gamma'_1 Z_{i,t-1} + \Gamma'_2 Y_{t-1}  + \tilde{e}_{i,t}   \label{eq:indicator_heterogeneous_average}
\end{align}
for $\bar{c} \in \{0,1\}$, which are variations of our baseline empirical specifications. Tables \ref{tab:appendix_indicator}  and \ref{tab:appendix_indicator_average} in Appendix \ref{sec:appendix_additional_firm_level_results} show the results under these alternative empirical specifications, with results providing support of our main findings. 

We also test whether our results are robust to our measure of leverage. Our empirical specifications use leverage measured as the ratio of total firm liabilities to total assets. In particular, we test whether results still arise when we use alternative measures of of liabilities. In particular, we consider three different measures of liabilities: (i) ratio of short-term or under a year debt to total assets, (ii) ratio of long-term or above a year debt to total assets, (iii) ratio of debt with banks and/or other domestic financial institutions to total assets. These different leverage specifications are motivated by the differences in borrowing patterns between Emerging Market and US companies. While US companies tend to primarily borrow long term, through commercial papers and not through banks, Emerging Market economies' companies tend to borrow at lower maturities, relying relatively more in domestic banks to fund working capital and capital expenditures.

While the magnitude of heterogeneity differs across specifications our benchmark results are robust across the different leverage measures. Results are presented in Tables \ref{tab:baseline_heterogeneous_def2} through \ref{tab:baseline_heterogeneous_average_def4} in Appendix \ref{sec:appendix_additional_firm_level_results}. To begin with, results are larger in magnitude when we consider long-term debt, proxied by liabilities or debt issued with a maturity greater than one year, compared to the short-term specification. Second, results are quantitatively smaller and less statistically significant for the leverage measure constructed using debt with banks or domestic financial institutions. Consequently, our benchmark results are robust given that short-term debt is usually associated with the financing of working capital and long-term debt is usually associated with capital expenditures.
 
Finally, we show that results are robust to different specifications of the exogenous US interest rate shock considered. The US interest rate shocks constructed by \cite{bauer2022reassessment} explicitly control for any information disclosed around FOMC meetings (see \cite{jarocinski2020deconstructing}) which matter for the transmission of US interest rate shocks for Emerging Markets (see \cite{camara2021spillovers}). However, to stay close to the empirical specifications of \cite{ottonello2020financial}, we test whether our results are robust to using a similar approach which relies only on the high-frequency movement of the Fed Fund Futures around FOMC meetings. We denote this high-frequency surprise by $\epsilon^{FFF}_t$ constructed as
\begin{align} \label{eq:shock_construction}
    \epsilon^{FFF}_t = FFF_{t+\Delta_{+}}-FFF_{t+\Delta_{-}}
\end{align}
where $t$ is the time of the monetary announcement, $FFF_t$ is the Federal Funds Futures contract for the current month at time $t$, $\Delta_{+}$ and $\Delta_{-}$ control the size of the time window around the announcement. We follow \cite{nakamura2018high} and use a time window of 30 minutes, from 10 minutes before the FOMC announcement ($\Delta_{-}=10$ minutes) to 20 minutes after it ($\Delta_{+}=20$ minutes). This high-frequency shocks are sourced from \cite{acosta2020estimating}.\footnote{In particular, the series of shocks can be downloaded from \url{https://www.acostamiguel.com/research.html}.}  Tables \ref{tab:baseline_heterogeneous_FFF} and \ref{tab:baseline_heterogeneous_average_FFF} show that results are robust for both the empirical specifications presented in Equations \ref{eq:baseline_heterogeneous} and \ref{eq:baseline_heterogeneous_average}, respectively.  

%%%%%%%%%%%%%%%%%%%%%%%%%%%%%%%%%%%%%%%%%%%%%%%%%%%%%%%%%%%%%%%%%%%%%%
%% Stylized Model
\section{Stylized Model of Tightening Borrowing Constraints} \label{sec:stylized_model}

In this section of the paper, we construct a stylized model that rationalize our firm level empirical findings. In our model, entrepreneurs consume and produce using capital, can borrow from international capital markets and face a standard leverage constraint. In a two period model without uncertainty we show that unconstrained agents' capital accumulation reacts stronger than constrained agents in response to an international interest rate movements. However, we show that in the same framework a tighter leverage constraint leads to a sharp reduction in capital accumulation of constrained agents while not affecting unconstrained agents.

The economy is populated by a continuum of mass 1 of heterogeneous entrepreneurs which live two periods indexed by $t=\{0,1\}$. Entrepreneurs value consumption in both periods and their preferences are represented by the following utility function
\begin{align}
    \mathcal{U} = \ln c_{i,0} + \beta \ln c_{i,1}
\end{align}
where $c_{i,t}$ represents consumption for periods $t=0$ and $t=1$, respectively. 

Entrepreneurs have access to a concave production technology function which uses only capital as input
\begin{align}
    y_{i,t} = k^{\alpha}_{i,t}
\end{align}
where $y_{i,t}$ represents quantities produced, $k_{i,t}$ represents capital used for production in period $t$, which has been installed in period $t-1$, and $\alpha \in \left(0,1\right)$ is a parameter which governs the concavity of the production function. 

Entrepreneurs have access to international capital markets and can borrow, denoted by $b_t$, at an interest rate of $r_t$. Entrepreneurs face a standard leverage constraint which implies that they can borrow up to a fraction of the value of their capital
\begin{align} \label{eq:leverage_constraint}
    b_{i,1} \leq \theta k_{i,1}
\end{align}
where $b_{i,1}$ and $k_{i,1}$ denote the stock of debt and capital entrepreneur $i$ chooses in period $0$ and carries into period $1$.

We can express entrepreneur's budget constraints for periods $t=0$ and $t=1$ as
\begin{align}
    c_{i,0} + k_{i,1} &= k^{\alpha}_{i,0} + b_{i,1} - b_{i,0} \left(1+r_0\right) \\
    c_{i,1} &= k^{\alpha}_{i,1} - b_{i,1} \left(1+r_1\right)
\end{align}
respectively, where $k_{i,0}$ and $b_{i,0}$ are given at the beginning of period 0. We assume that the only dimension of heterogeneity across entrepreneurs is their initial debt holdings: $b_{i,0}$. This implies that entrepreneurs differ in their initial leverage.

We can state entrepreneur $i$'s utility maximization problem as 
\begin{equation*}
    \begin{aligned}
    \max_{c_{i,0},c_{i,1},k_{i,1},b_{i,1}} \quad &  \mathcal{U} = \ln c_{i,0} + \beta \ln c_{i,1} \\
    \textrm{s.t.} \quad & c_{i,0} + k_{i,1} = k^{\alpha}_{i,0} + b_{i,1} - b_{i,0} \left(1+r_0\right) \\
    & c_{i,1} = k^{\alpha}_{i,1} - b_{i,1} \left(1+r_1\right) \\
      & b_{i,1} \leq \theta_i k_{i,1}
    \end{aligned}
\end{equation*}
The entrepreneur utility maximization problem's solution can be characterized by the Euler equations for capital $k_{i,1}$ and debt $b_{i,1}$
\begin{align}
    \frac{1}{c_{i,0}} &= \frac{\beta}{c_{i,1}} \alpha k^{\alpha-1}_{i,1} + \mu \theta \label{eq:optimal_capital_euler} \\
    \frac{1}{c_{i,0}} &= \frac{\beta}{c_{i,1}} \left(1+r_1\right) + \mu \label{eq:optimal_bond_euler}
\end{align}
where $\mu$ represents the Lagrange multiplier on the leverage constraint in Equation \ref{eq:leverage_constraint}. On the one hand, if the leverage constraint does not bind, i.e. $\mu$ equals zero, the optimal stock of capital can be computed by expressing the Euler equations without the Lagrange multiplier
\begin{align*}
    \frac{1}{c_{i,0}} &= \frac{\beta}{c_{i,1}} \alpha k^{\alpha-1}_{i,1} \\
    \frac{1}{c_{i,0}} &= \frac{\beta}{c_{i,1}} \left(1+r_1\right)
\end{align*}
Some algebraic manipulation leads to the following expression for the optimal stock of capital
\begin{align} \label{eq:optimal_capital_accumulation_unconstrained}
    k^{*}_{i,1} = \left( \frac{1+r_1}{\alpha} \right)^{\frac{1}{\alpha-1}}
\end{align}
which is a decreasing function of the gross interest rate $1+r_1$.

On the other hand, if the leverage constraint binds, i.e. $\mu > 0$, the optimal choice of capital can be characterized by the following expression
\begin{align} \label{eq:optimal_capital_constraint}
    \alpha k^{\alpha-1} - \left(1+r_1\right) = \mu \frac{c_1}{\beta} \left(1-\theta\right)
\end{align}
which emerges from combining Equations \ref{eq:optimal_capital_euler} and \ref{eq:optimal_bond_euler}. Given that the Lagrange multiplier on the leverage constraint $\mu$ is positive and that parameter $\theta$ in the entrepreneur's leverage constraint is strictly lower than 1, this implies that the marginal product of capital is greater than the gross interest rate. This implies that the constrained entrepreneur's capital accumulated to period $1$ is lower than that of unconstrained entrepreneurs. 

Next, we show both analytically and graphically that unconstrained entrepreneurs exhibit a greater response to changes in the gross interest rate, $1+r_1$, than constrained entrepreneurs. Proposition 1 shows that for close enough initial conditions, the response of entrepreneur's optimal capital accumulation, $k_{1}$, to the interest rate, $1+r_1$, is greater for unconstrained entrepreneurs than for constrained entrepreneurs. 

\noindent
\textbf{Proposition 1:} For a given set of parameters $\{\alpha,\theta\}$, initial capital stock $k_0$, and close enough initial stock of debt $b_0$ such that one entrepreneur is unconstrained and a second one is constrained, the response of the optimal stock of capital $k_1$ with respect to the gross interest rate $r_1$ is greater in magnitude for the unconstrained entrepreneur than for the constrained entrepreneur.

\noindent
\textit{See Appendix \ref{sec:appendix_additional_structural_model_details} for proof.}

\noindent
In order to understand the intuition behind this result it is useful to think about two entrepreneur, one unconstrained and one constrained, i.e. borrowing up to its limit. On the one hand, if the former entrepreneur remains unconstrained, Equation \ref{eq:optimal_capital_accumulation_unconstrained} implies that the entrepreneur will increase its optimal capital accumulation. On the other hand, the entrepreneur which is already at the borrowing constraint would like to increase its capital accumulation but cannot access additional resources due to low initial endowments, $k^{\alpha}_0 - b_0 \left(1+r_0\right)$, and the binding borrowing constraint. Consequently, the magnitude of the change in the optimal capital accumulation $k^{*}_1$ will be greater for the unconstrained entrepreneur compared to that of the constrained entrepreneur.

\begin{figure}[ht]
    \centering
    \caption{Optimal Capital Accumulation \& Interest Rate}
    \label{fig:Model_Capital_Accumulation_Interest_Rate_Smaller}    
    \includegraphics[scale=0.65]{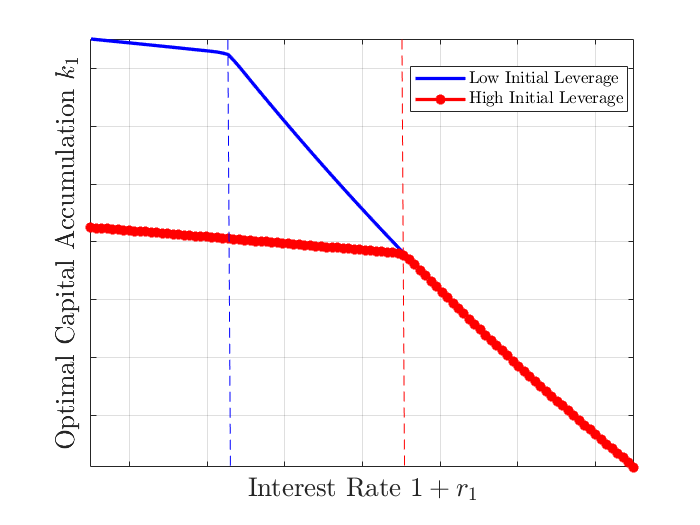}
    \floatfoot{\textbf{Note:} The figure above represents the entrepreneur's optimal capital accumulation $k^{*}_1$ (y-axis) as a function of the interest rate $(1+r_1)$ (x-axis). The blue line represents an entrepreneur with relatively low initial leverage while the dotted red line represents an entrepreneur with relatively high initial leverage. This is computed as entrepreneurs having the same initial capital stock, stock $k_0$, but the higher leverage entrepreneur having a greater initial stock of debt, $b_0$. Starting from the right, the blue and red dashed lines represents the interest rate at which entrepreneurs stop being at the borrowing constraint.}
\end{figure}

Figure \ref{fig:Model_Capital_Accumulation_Interest_Rate_Smaller} presents this result graphically.
The figure plots the optimal capital accumulation, $k^{*}_1$, as a function of the interest rate, $(1+r_1)$, for two entrepreneurs, one with a low initial leverage (blue solid line) and one with a high initial leverage (red dotted line). First, starting from the right hand side of the figure, the optimal capital accumulation exhibits a kink at the interest rate in which entrepreneurs become borrowing constrained. In this simple economy, it is clear that the slope of this figures is significantly smaller in magnitude to the left of the kink. Second, as long as both entrepreneurs are not at the borrowing constraint (i.e. to the right of the dashed red line or right-most vertical dashed line), their optimal capital accumulation $k_1$ is identical.\footnote{Note that this can also be observed from noting that the expression for the unconstrained optimal capital accumulation in Equation \ref{eq:optimal_capital_accumulation_unconstrained} does not depend on an entrepreneur's initial conditions such as $k_0$, $b_0$.} Moving towards the left of the figure, as the ``high initial leverage'' entrepreneur becomes constrained but the ``low initial leverage'' entrepreneur remains unconstrained, the capital accumulation of later becomes increasingly greater than that of the latter.

\begin{figure}[ht]
    \centering
    \caption{Optimal Capital Accumulation \& Interest Rate}
    \label{fig:Model_Capital_Accumulation_Interest_Rate_Smaller_Arrows}    
    \includegraphics[scale=0.65]{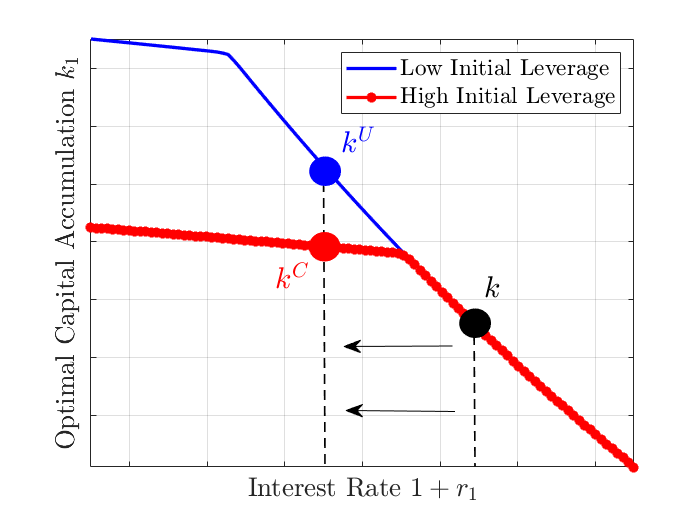}
    \floatfoot{\textbf{Note:} The figure above represents the entrepreneur's optimal capital accumulation $k^{*}_1$ (y-axis) as a function of the interest rate $(1+r_1)$ (x-axis). The blue line represents an entrepreneur with relatively low initial leverage while the dotted red line represents an entrepreneur with relatively high initial leverage. This is computed as entrepreneurs having the same initial capital stock, stock $k_0$, but the higher leverage entrepreneur having a greater initial stock of debt, $b_0$. Starting from the right, the blue and red dashed lines represents the interest rate at which entrepreneurs stop being at the borrowing constraint.}
\end{figure}

Figure \ref{fig:Model_Capital_Accumulation_Interest_Rate_Smaller_Arrows} shows graphically how under this framework a decrease in the interest rate leads to a greater capital accumulation for the unconstrained entrepreneur, from $k$ to $k^{U}$, than for the constrained entrepreneur, from $k$ to $k^{C}$. 
Furthermore, the fact that unconstrained firms have a greater response in their investment to changes in interest rates than constrained firms has being empirically corroborated. This result is at the core of \cite{ottonello2020financial}, which finds that US firms with a greater distance to default (and those which are relatively less indebted) exhibit a greater response of investment than firms closer to default (and those which are relatively more indebted) after a US interest rate shock.

We show analytically and graphically that a greater response of capital accumulation from constrained entrepreneurs can be achieved through a change in financial conditions, triggered by a looser borrowing constraint or greater $\theta$. Proposition 2, shows this under our present stylized model. 

\noindent
\textbf{Proposition 2:} For a given parameter $\alpha$ and initial conditions $\{k_0,b_0\}$, the constrained entrepreneur's optimal capital accumulation, $k_1$, is an increasing function of parameter $\theta$, provided the entrepreneur remains borrowing constrained. 

\noindent
\textit{See Appendix \ref{sec:appendix_additional_structural_model_details} for proof.}

\begin{figure}[ht]
    \centering
    \caption{Optimal Capital Accumulation \& Financial Conditions}
    \label{fig:Model_Capital_Accumulation_Initial_Debt_Smaller_Dashed}    
    \includegraphics[scale=0.65]{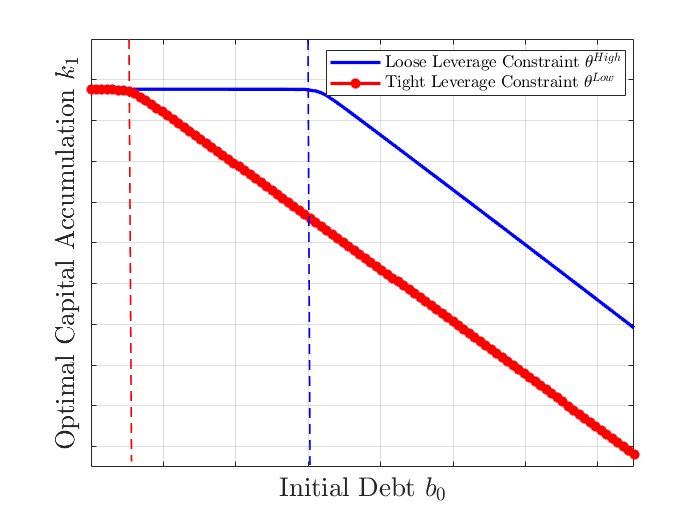}
    \floatfoot{\textbf{Note:} The figure above represents the entrepreneur's optimal capital accumulation $k^{*}_1$ (y-axis) as a function of their initial debt $b_0$ (x-axis). The blue line represents an entrepreneur that faces a relatively loose borrowing constraint, i.e. a borrowing constraint with a greater $\theta$, while the dotted red line represents an entrepreneur which faces a relatively tighter borrowing constraint, i.e. a borrowing constraint with a lower $\theta$.}
\end{figure}

Figure \ref{fig:Model_Capital_Accumulation_Initial_Debt_Smaller_Dashed} shows the intuition behind this result by plotting the relationship between optimal capital accumulation, $k_1$, and initial debt, $b_0$, for entrepreneurs who face differently tight borrowing constraints.
The solid blue line represents the optimal capital accumulation for an entrepreneur which faces a relatively looser borrowing constraint, i.e. a borrowing constraint with a greater $\theta$. The dotted red line represents the optimal capital accumulation for an entrepreneur which faces a relatively tighter borrowing constraint, i.e., a borrowing constraint with a smaller $\theta$. Starting from the left, on the segment of the figure to the left of the vertical dashed red line (to the left of the left-est vertical dashed line) the optimal capital accumulation, $k_1$, is identical for both entrepreneurs. However, as we move to the right, even for the same initial debt, $b_0$, entrepreneurs which face a tighter borrowing constraint exhibit lower capital accumulation (dotted red line) than firms which face a looser borrowing constraint (solid blue line). 

\begin{figure}[ht]
    \centering
    \caption{Optimal Capital Accumulation \& Financial Conditions}
    \label{fig:Model_Capital_Accumulation_Initial_Debt_Smaller_Dashed_Arrows}    
    \includegraphics[scale=0.65]{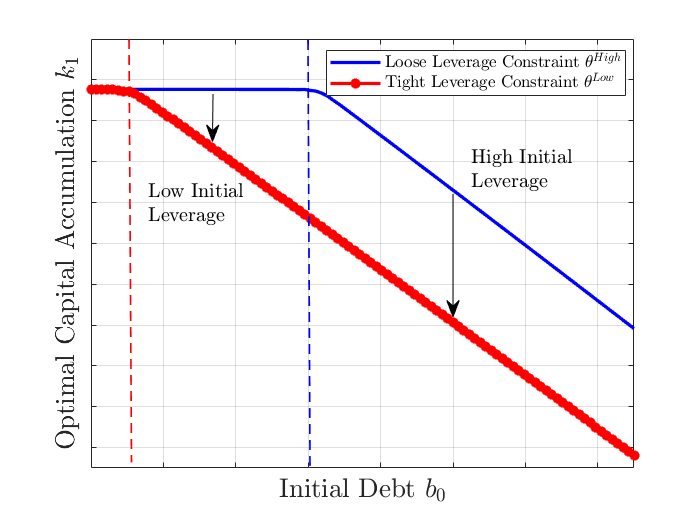}
    \floatfoot{\textbf{Note:} The figure above represents the entrepreneur's optimal capital accumulation $k^{*}_1$ (y-axis) as a function of their initial debt $b_0$ (x-axis). The blue line represents an entrepreneur that faces a relatively loose borrowing constraint, i.e. a borrowing constraint with a greater $\theta$, while the dotted red line represents an entrepreneur which faces a relatively tighter borrowing constraint, i.e. a borrowing constraint with a lower $\theta$.}
\end{figure}

Figure \ref{fig:Model_Capital_Accumulation_Initial_Debt_Smaller_Dashed_Arrows} shows how an entrepreneur which have a greater in initial leverage, proxied by different levels of initial debt $b_0$, exhibit a greater drop in capital accumulation than entrepreneurs with lower initial leverage. 
The figure shows that, in response to a reduction in $\theta$, the drop in capital accumulation is greater for an entrepreneur with a high initial level of debt (to the right of the dashed blue line) than for an entrepreneur with low initial debt (to the left of the dashed blue line). For an initially constrained entrepreneur, the drop in $\theta$ leads to a one-to-one reduction in leverage which is achieved through a drop in capital accumulation. If the entrepreneur is initially unconstrained and remains so after the drop in $\theta$, its level of capital accumulation will not be affected (to the left of the dashed red line in Figure \ref{fig:Model_Capital_Accumulation_Initial_Debt_Smaller_Dashed_Arrows}). For an initially constrained entrepreneur which becomes constrained due to the reduction in $\theta$, the drop in capital accumulation will be somewhere in between the two extreme cases, depending on how close to the borrowing constraint it was before the shock. 

In summary, higher borrowing rates and higher borrowing constraints affect firms differently according to their initial leverage. On the one hand, firms relatively far from their leverage constraints respond greatly to higher interest rates while not responding to tighter borrowing constraints. On the other hand, firms relatively close to their borrowing constraints do not respond significantly to changes in borrowing constraints, but are forced to cut drastically on their investment when leverage constraints become tighter. In light of this stylized model's predictions, our firm level findings presented in Section \ref{sec:firm_evidence} suggest that the international transmission of US monetary policy shocks on firms investment is primarily through tighter borrowing conditions and not necessarily higher borrowing rates. 

\section{Conclusion} \label{sec:conclusion}

In this paper, we argued that US monetary policy shocks are transmitted into Emerging Markets via tighter borrowing constraints which reduces investment. First, we showed that for a panel of Emerging Market economies, a US monetary policy shock is associated with an economic recession driven by a persistent drop of aggregate investment. Second, exploiting more than a decade of balance sheet data of Chilean firms, we showed that relatively more indebted firms experience a larger investment drop than relatively less indebted firms after a US monetary policy shock. This result stands out as it is completely at odds with the findings for US firms, even when using the same econometric approach and identification strategy. 

We interpret our results through the lenses of a stylized model of heterogeneous entrepreneurs. Entrepreneurs have access to production function which uses capital only and can borrow up to a fraction of the value of their capital. Under this framework, unconstrained entrepreneurs equate the marginal product of their capital with the borrowing interest rate. Consequently, their capital accumulation responds significantly to changes in the borrowing rates. Constrained entrepreneurs do not react significantly to changes in borrowing rates but must cutback drastically in response to tighter financial conditions. Thus, we argue that our firm level results suggest that US monetary policy shocks are associated with international tighter borrowing conditions that go beyond higher borrowing rates.

Overall, our results suggests that the investment channel plays a crucial role on the international transmission of US interest rate shocks. Furthermore, our firm level results suggest that the transmission of US interest rate shocks is different within than outside the US economy.

%%%%%%%%%%%%%%%%%%%%%%%%%%%%%%%%%%%%%%%%%%%%%%%%%%%%%%%%%%%%%%%%%%%%%%%%%%%%%%%%%%%%%%
\newpage
\bibliography{main.bib}

%%%%%%%%
%%%%%%%%
\newpage
\appendix

%%%%%%%%%%%%%%%%%%%%%%%%%%%%%%%%%%%%%%%%%%%%%%%%%%%%%%%%%%%%%%%%%%%%%%%
\newpage
\section{Country Level Data Description} \label{sec:appendix_aggregate_data}

In this appendix we present additional description on the datasets used across the paper.
\begin{table}[ht]
    \centering
    \caption{Country List}
    \label{tab:appendix_country_list}
    \begin{tabular}{c}
       Brazil        \\
       Chile         \\
       Indonesia     \\
       Mexico        \\
       Peru          \\
       South Africa  \\
       %Thailand      \\
       Turkey        \\
    \end{tabular}
\end{table}
In particular, Table \ref{tab:appendix_country_list} describes the list of countries included in the empirical exercises across the paper. The main empirical results presented in this paper are based in 8 Emerging Market economies: Brazil, Chile, Indonesia, Mexico, Peru, South Africa and Turkey for the period January-2004 to December 2018. The choice of time-window is motivated by our firm-level data availability for Chile. 

%%%%%%%%%%%%%%%%%%%%%%%%%%%%%%%%%%%%%%%%%%%%%%%%%%%%%%%%%%%%%%%%%%%%%%%
\newpage
\section{SVAR Model Details} \label{sec:appendix_model_details}

In this section of the appendix I provide additional details on the estimation of the Structural VAR models presented in Section \ref{sec:aggregate_evidence}. 

First, we describe the panel SVAR model used to estimate the results presented in Figure \ref{fig:Only_Domestic_Panel_Larger_Norm} in Section \ref{sec:aggregate_evidence}. The model described by Equation \ref{eq:pooled_estimator} is estimated using Bayesian methods. In order to carry out the estimation of this model I first re-write the model. In particular, the model can be reformulated in compact form as

\begin{align} \label{eq:model_compact_form}
\underbrace{\begin{pmatrix}
y_{1,t}' \\
y_{2,t}' \\
\vdots  \\
y_{N,t}'
\end{pmatrix}}_{Y_t, \quad N \times n}
&=
\underbrace{\begin{pmatrix}
y_{1,t-1}' \ldots y_{1,t-p}' \\
y_{2,t-1}' \ldots y_{2,t-p}' \\
\vdots  \ddots \vdots \\
y_{N,t-1}' \ldots y_{N,t-p}'
\end{pmatrix}}_{\mathcal{B}, \quad N \times np}
\underbrace{\begin{pmatrix}
\left(A^{1}\right)' \\
\left(A^{2}\right)' \\
\vdots \\
\left(A^{N}\right)'
\end{pmatrix}}_{X_t, \quad np \times n}
+
\underbrace{\begin{pmatrix}
\epsilon_{1,t}' \\
\epsilon_{2,t}' \\
\vdots \\
\epsilon_{N,t}'
\end{pmatrix}}_{\mathcal{E}_t, \quad N \times n}
\end{align}
or
\begin{align}
    Y_t = X_t \mathcal{B} + \mathcal{E}_t
\end{align}
Even more, the model can be written in vectorised form by stacking over the $T$ time periods 
\begin{align}
    \underbrace{vec\left(Y\right)}_{NnT \times 1} = \underbrace{\left(I_n \otimes X \right)}_{NnT \times n np} \quad \underbrace{vec\left(\mathcal{B}\right)}_{n np \times 1} \quad + \quad \underbrace{vec\left(\mathcal{E}\right)}_{NnT \times 1}
\end{align}
or
\begin{align}
    y = \Bar{X} \beta + \epsilon
\end{align}
where $\epsilon \sim \mathcal{N}\left(0, \Bar{\Sigma}\right)$, with $\Bar{\Sigma} = \Sigma_c \otimes I_{NT}$.

The model described above is just a conventional VAR model. Thus, the traditional Normal-Wishart identification strategy is carried out to estimate it. The likelihood function is given by
\begin{align}
    f\left(y | \Bar{X} \right) \propto |\Bar{\Sigma}|^{-\frac{1}{2}} \exp \left(-\frac{1}{2} \left(y - \Bar{X}\beta\right)' \Bar{\Sigma}^{-1} \left(y - \Bar{X}\beta\right) \right)
\end{align}
As for the Normal-Wishart, the prior of $\beta$ is assumed to be multivariate normal and the prior for $\Sigma_c$ is inverse Wishart. For further details, see \cite{dieppe2016bear}.

Second, we describe the single country SVAR model used to estimate the impact of US monetary policy shocks for the Chilean economy. 

%%%%%%%%%%%%%%%%%%%%%%%%%%%%%%%%%%%%%%%%%%%%%%%%%%%%%%%%%%%%%%%%%%%%%%%
\newpage
\section{Additional SVAR Model Results} \label{sec:appendix_model_svar_results}

In this Appendix we present additional results which complement and provide robustness to those presented in Section \ref{sec:aggregate_evidence}. Section \ref{subsec:appendix_model_svar_results_panel} presents additional results for a panel of countries using a panel SVAR model. Section \ref{subsec:appendix_model_svar_results_chile} presents additional results for the Chilean economy using a single country SVAR model. 

\subsection{Additional SVAR Model Results: Panel} \label{subsec:appendix_model_svar_results_panel}

In this Appendix we present additional results using a panel SVAR model. The model follows the same specifications as those presented in Section \ref{sec:aggregate_evidence} and described in greater detail in Appendix \ref{sec:appendix_additional_structural_model_details}.

\begin{figure}[ht]
    \centering
    \caption{IRF Analysis - US Tightening Shock \\ \footnotesize Panel SVAR -  All Variables}
    \label{fig:Full_Panel_Larger_Norm}    
    \includegraphics[width=14cm,height=10cm]{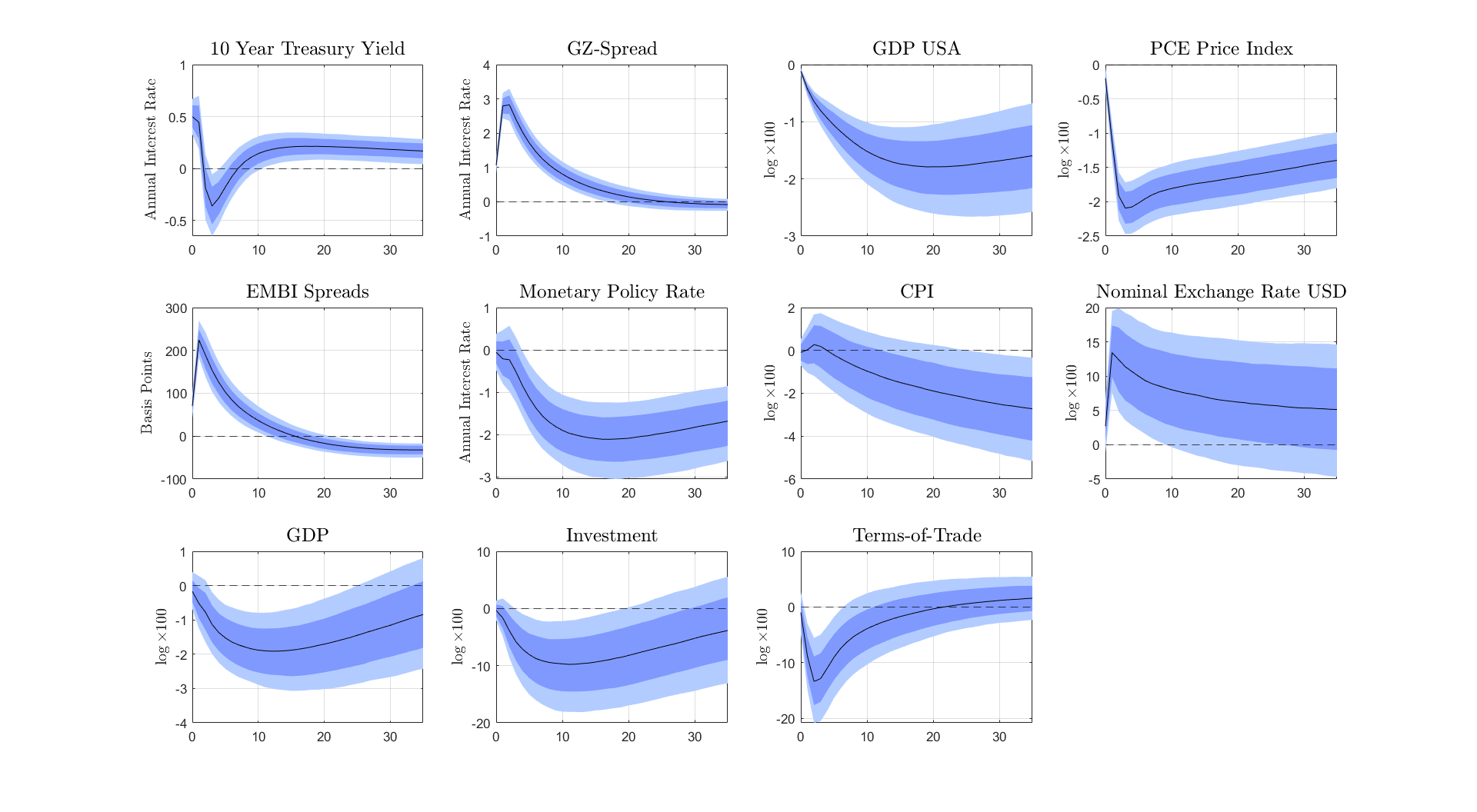}
    \floatfoot{\textbf{Note}: Median (black line), (68\% confidence interval in dark blue band), (90\% confidence interval in light blue band). Panels' order starts from the top left corner, moving from left to right, and later from top to bottom. The impulse response functions are computed from a exogenous US tightening shock which causes the 10-Year Treasury yield increase by 50 basis points.}
\end{figure}

First, Figure \ref{fig:Full_Panel_Larger_Norm} complements the results presented in Figure \ref{fig:Only_Domestic_Panel_Larger_Norm} by showing the dynamics of all variables in the model, i.e. USA's variables and domestic variables. 
As stated, the US' 10-Year Treasury yield increases by 50 basis points initially, see Panel (1.1). While this initial increase is followed by sharp drop in the following period, the yield remains above its pre-crisis level even 36 months after the initial shock. The monetary policy shock leads to a hump-shaped and persistent increase in the USA's private interest rates, proxied by \cite{gilchrist2012credit} spreads, see in Panel (1.2). USA's domestic GDP and PCE index exhibit a sharp and persistent drop. Interestingly, the drop in GDP is larger for the Emerging Markets than in the USA. The results are in line with a previous literature such as \cite{bauer2022reassessment}. 

Second, we carry out a robustness check by substituting the real GDP index by an industrial production index in the panel SVAR sample. In Section \ref{sec:aggregate_evidence} we introduced a real GDP index in our benchmark panel SVAR specification as our main goal was to gauge the quantitative role of the investment channel in explaining the economic recession that followed a US monetary policy shock. However, as stated in Section \ref{sec:aggregate_evidence}, the real GDP index for several countries in our sample is constructed as an interpolation of quarterly data. 
\begin{figure}[ht]
    \centering
    \caption{IRF Analysis - US Tightening Shock \\ \footnotesize Panel SVAR -  Industrial Production }
    \label{fig:Full_Panel_Larger_Ind_Norm}    
    \includegraphics[width=14cm,height=10cm]{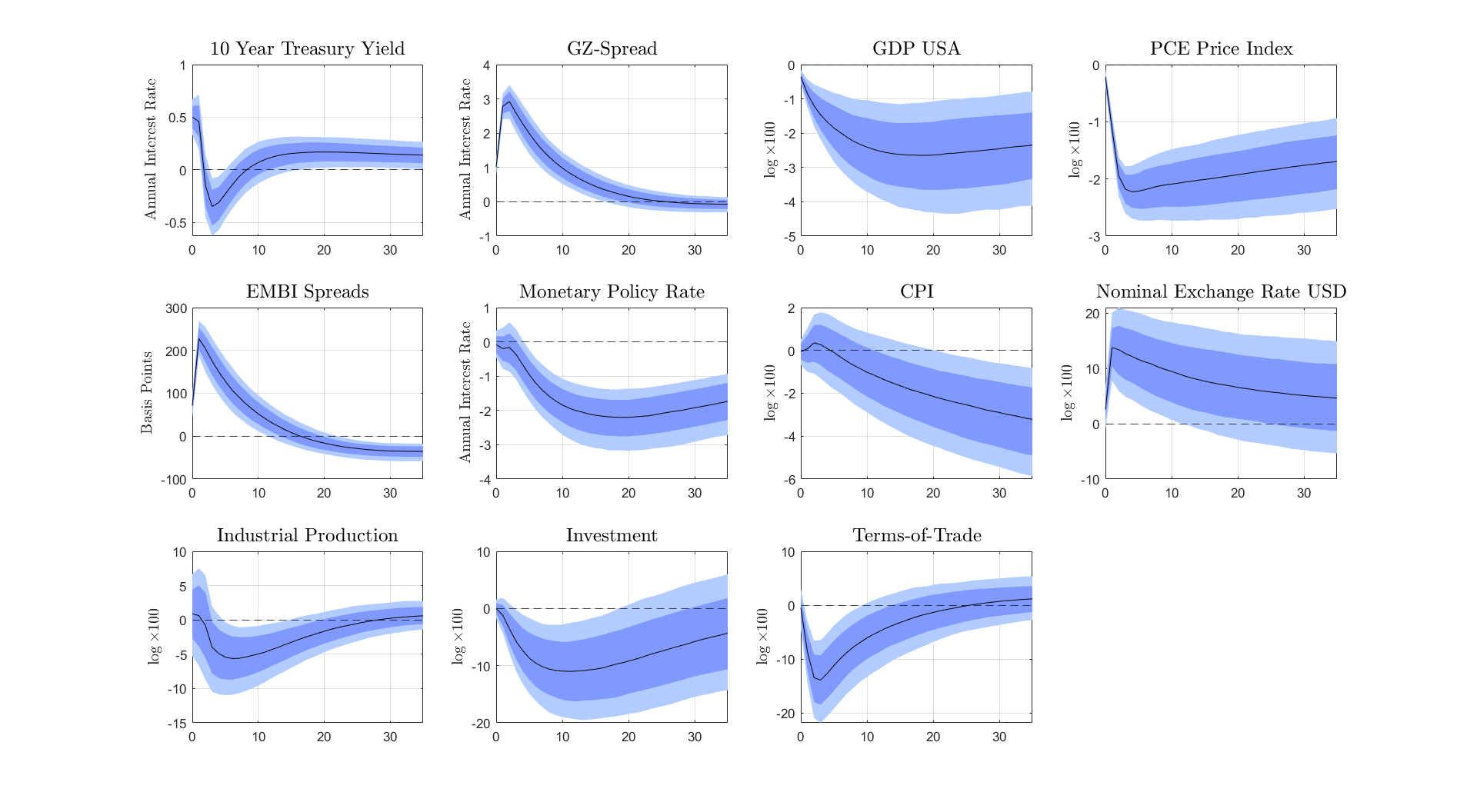}
    \floatfoot{\textbf{Note}: Median (black line), (68\% confidence interval in dark blue band), (90\% confidence interval in light blue band). Panels' order starts from the top left corner, moving from left to right, and later from top to bottom. The impulse response functions are computed from a exogenous US tightening shock which causes the 10-Year Treasury yield increase by 50 basis points.}
\end{figure}
In order to deal with any potential issues arising from this interpolation, we substitute the real GDP index with a monthly constructed industrial production index sourced from the IMF IFS dataset. Figure \ref{fig:Full_Panel_Larger_Ind_Norm} shows that our benchmark results are robust to this robustness check. The industrial production index exhibits a large drop, reaching a drop close to 5\% below its pre-shock levels. For comparison, the drop in the GDP index was close to 2.5\%. Dynamics for other variables are roughly close to those presented in Figures \ref{fig:Only_Domestic_Panel_Larger_Norm} and \ref{fig:Full_Panel_Larger_Norm}.

\subsection{Additional SVAR Model Results: Chile} \label{subsec:appendix_model_svar_results_chile}

In this Appendix we present additional results for the Chilean economy using a single country SVAR model. The model follows the same specifications as those presented in Section \ref{sec:aggregate_evidence} and described in greater detail in Appendix \ref{sec:appendix_additional_structural_model_details}.

To begin with, we start by introducing the dynamics of both USA's and Chile's variables, i.e. the all variables in our benchmark SVAR specification. 
\begin{figure}[ht]
    \centering
    \caption{IRF Analysis - US Tightening Shock \\ \footnotesize Chile SVAR -  All Variables }
    \label{fig:Chile_Full_Larger}    
    \includegraphics[width=14cm,height=10cm]{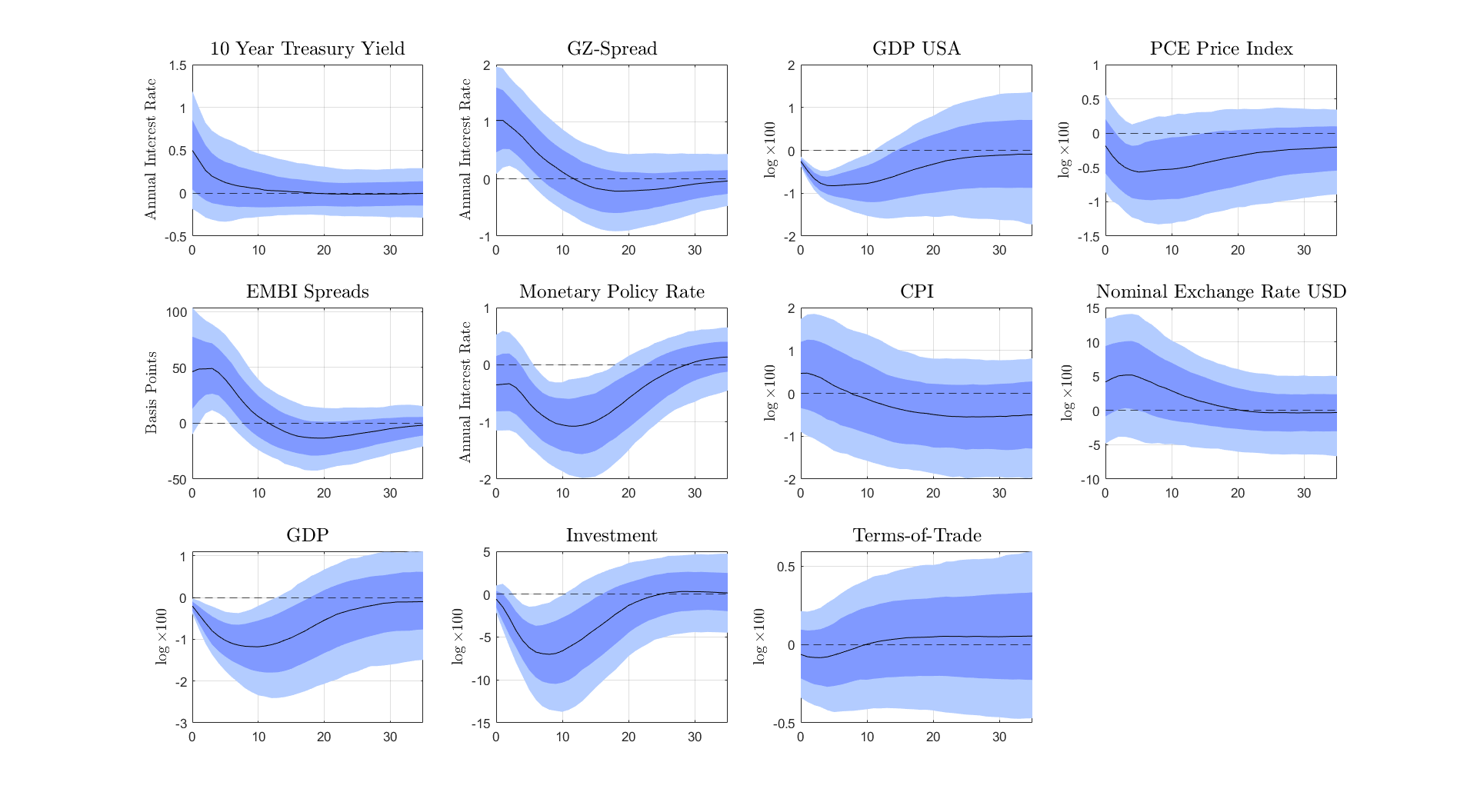}
    \floatfoot{\textbf{Note}: Median (black line), (68\% confidence interval in dark blue band), (90\% confidence interval in light blue band). Panels' order starts from the top left corner, moving from left to right, and later from top to bottom. The impulse response functions are computed from a exogenous US tightening shock which causes the 10-Year Treasury yield increase by 50 basis points.}
\end{figure}
The dynamics of USA's variables exhibited in Panels (1.1) through (1.4) are in line with the results presented in Figure \ref{fig:Full_Panel_Larger_Norm} for the panel SVAR model. 

Next, we carry out a robustness check by replacing the GDP index by an industrial production index. 
\begin{figure}[ht]
    \centering
    \caption{IRF Analysis - US Tightening Shock \\ \footnotesize Chile -  Industrial Production}
    \label{fig:Chile_Full_Larger_Ind}    
    \includegraphics[width=14cm,height=12cm]{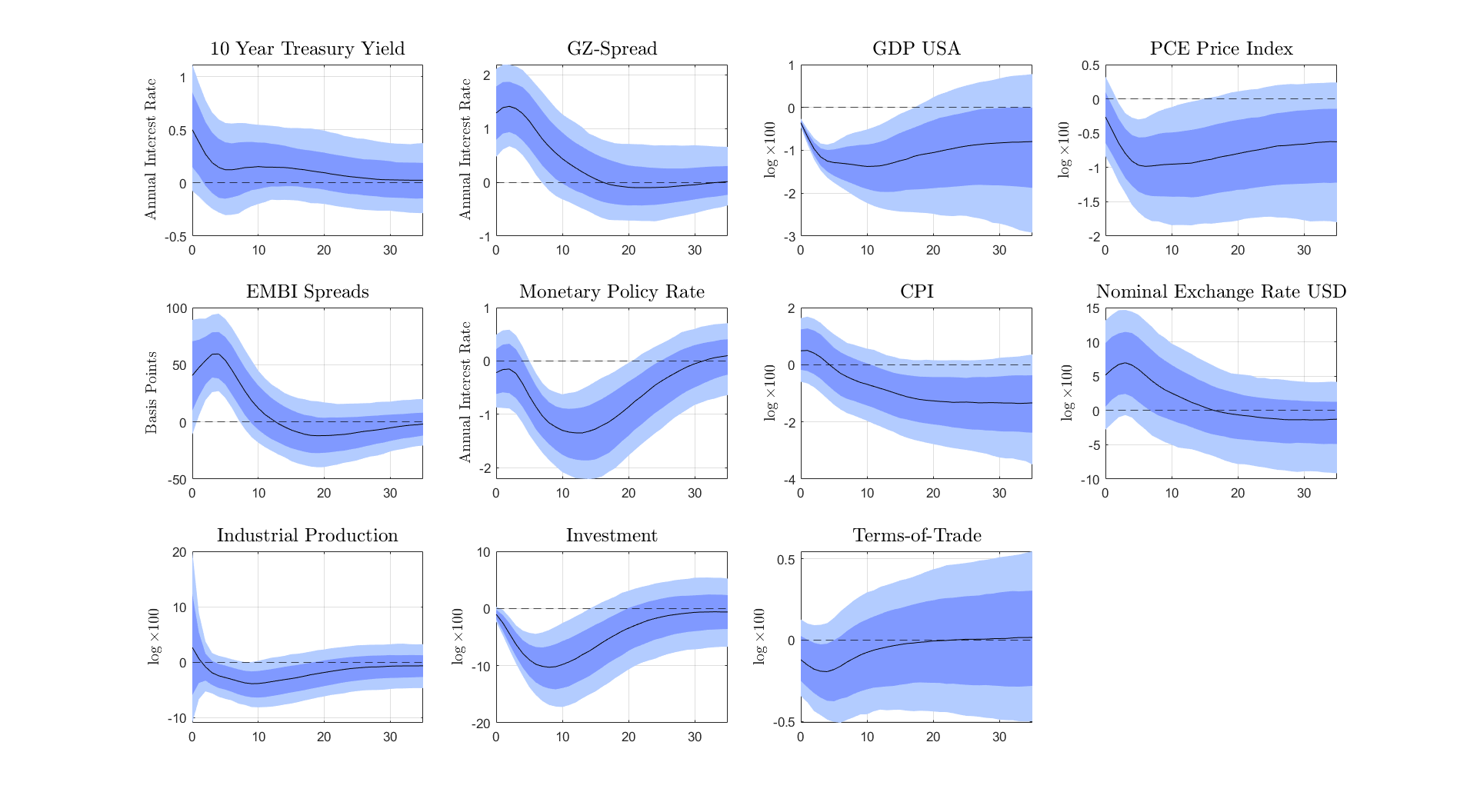}
    \floatfoot{\textbf{Note}: Median (black line), (68\% confidence interval in dark blue band), (90\% confidence interval in light blue band). Panels' order starts from the top left corner, moving from left to right, and later from top to bottom. The impulse response functions are computed from a exogenous US tightening shock which causes the 10-Year Treasury yield increase by 50 basis points.}
\end{figure}
Note that the Chilean Central Bank produces a monthly GDP index called the ``IMACEC'' index.\footnote{See \url{https://www.bcentral.cl/web/banco-central/areas/estadisticas/imacec}} Consequently, a priori, the results presented in Figure \ref{fig:Chile_OnlyDomestic_Larger} is not in risk of biases introduced by interpolation. However, as other work in the literature use industrial production (see \cite{camara2021spillovers}), and we carry out this robustness check for our panel SVAR specification (see Figure \ref{fig:Full_Panel_Larger_Ind_Norm}), we also carry out this robustness check for the Chilean economy specification. Figure \ref{fig:Chile_Full_Larger_Ind} shows the results are robust. The industrial production index exhibits a large and persistent drop, but with larger uncertainty bounds than those for the benchmark specification. The dynamics of the rest of the variables are in line with the results presented in the benchmark specification in Figure \ref{fig:Chile_Full_Larger}.

%%%%%%%%%%%%%%%%%%%%%%%%%%%%%%%%%%%%%%%%%%%%%%%%%%%%%%%%%%%%%%%%%%%%%%
\newpage
\section{Additional Firm Details} \label{sec:appendix_data_firms}

In this section of the paper we present additional details on the firm level dataset used in Section \ref{sec:firm_evidence}. We construct our sample of firm level variables from the quarterly CMF (\textit{Comision para el Mercado Financiero}) portal, a panel of publicly listed Chilean firms.

First, Table \ref{tab:firms_by_sectors} presents the coverage of firms across their different reported sector of activity.
\begin{table}[ht] %% Table on Sectors
    \centering
    \footnotesize
    \caption{Number of Firms by Sectors}
    \label{tab:firms_by_sectors}
    \begin{tabular}{l c c}
\textbf{Sector}	&	\textbf{Total Sample}	&	\textbf{Year - 2007}	\\ \hline \hline
Food, beverages and tobacco	&	2,294	&	185	\\
Consumption Industries	&	1,240	&	115	\\
Construction	&	1,524	&	123	\\
Forestry	&	562	&	49	\\
Transport and maritime services	&	766	&	52	\\
Rail and road transport	&	400	&	25	\\
Mining	&	785	&	64	\\
Energy	&	256	&	12	\\
Telecommunications	&	600	&	70	\\
Sanitary services and gas	&	669	&	52	\\
Electricity (utilities)	&	1,529	&	117	\\
Health services	&	301	&	17	\\
Infrastructure concessionaires	&	782	&	54	\\
Recreation and educational services	&	966	&	58	\\ \hline
\textbf{Total}	&	\textbf{12,674}	&	\textbf{993}	\\  \hline \hline
    \end{tabular}
\end{table}
This table suggests that our sample exhibits a representative sample of the Chilean economy. For the period 2004-2018 our sample counts with 990 firms. This is greater than the 228 firms with publicly traded stocks in the Chile's ``\textit{Santiago Stock Exchange}''.

%%%%%%%%%%%%%%%%%%%%%%%%%%%%%%%%%%%%%%%%%%%%%%%%%%%%%%%%%%%%%%%%%%%%
\newpage
\section{Additional Firm Level Results} \label{sec:appendix_additional_firm_level_results}

In this Appendix of the paper we present additional, complementary firm level results on the transmission of US interest rate shocks. 

First, we start by replacing the interaction term
\begin{align*}
    \left(l_{i,t-1} - \mathbb{E}_{i} \left[ l_{t}\right] \right) \epsilon^{X}_t
\end{align*}
with two interaction terms constructed using indicator functions
\begin{align*}
    \mathbbm{1} \left[\frac{l_{i,t} - \mathbb{E}_i \left[l_t \right]}{\sigma_{l} \left( l_t\right)}  > 0 \right] & \times \epsilon^{X}_t \\
    \mathbbm{1} \left[\frac{l_{i,t} - \mathbb{E}_i \left[l_t \right]}{\sigma_{l} \left( l_t\right)}  > 1 \right] & \times \epsilon^{X}_t
\end{align*}
where the first specification is an indicator function that takes the value of 1 if the standardized present measure of firm leverage is above its sample mean and zero otherwise, and the second specification is an indicator function which takes the value of 1 if the standardized present measure of firm leverage is one standard deviation above its sample mean and zero otherwise. 

Table \ref{tab:appendix_indicator} presents the results of estimating Equation \ref{eq:baseline_heterogeneous}.
\begin{table}[ht]
    \centering
    \caption{Heterogeneous Response of Investment  to $\epsilon^{X}_t$ \\ \footnotesize Indicator Function }
    \scriptsize
    \label{tab:appendix_indicator}
\begin{tabular}{lcccc}
 & \multicolumn{4}{c}{Firm investment - $\Delta \log k_{i,t}$ } \\
  & (1) & (2) & (3) & (4) \\ \hline \hline
 &  &  &  &  \\
$\mathbbm{1} \left[\frac{l_{i,t} - \mathbb{E}_i \left[l_t \right]}{\sigma_{l} \left( l_t\right)}  > 0 \right] \epsilon^{X}_t$  & -0.432*** & -0.417** & -0.311 & -0.318 \\
 & (0.142) & (0.156) & (0.197) & (0.201) \\
 \\
$\mathbbm{1} \left[\frac{l_{i,t} - \mathbb{E}_i \left[l_t \right]}{\sigma_{l} \left( l_t\right)}  > 1 \right] \epsilon^{X}_t$ & -0.691* & -0.722* & -0.534 & -0.553 \\
 & (0.373) & (0.369) & (0.359) & (0.356) \\
 &  &  &  &  \\
Firm FE & yes & yes & yes & yes \\
Sector-Time FE & yes & yes & yes & yes \\
Control for Lagged Inv. & no & yes & yes & yes \\
Control for Lagged Currency Mismatch & no & no & yes & yes \\
Control for $l_{i,t-1}$ & no & no & no & yes \\
 &  &  &  &  \\
Observations & 12,646 & 12,473 & 11,967 & 11,869 \\  \hline \hline
\multicolumn{5}{c}{ Clustered standard errors in parentheses} \\
\multicolumn{5}{c}{ *** p$<$0.01, ** p$<$0.05, * p$<$0.1} \\
\end{tabular}
\floatfoot{\textbf{Note:} Results in each row represent results of different regression. The results presented are estimates of parameter $\beta$ of one of the following two specifications. The first row of results are computed by estimating the specification: $\log k_{i,t} - \log k_{i,t-1} = \alpha_i + \alpha_{s,t} + \beta \mathbbm{1} \left[\frac{l_{i,t} - \mathbb{E}_i \left[l_t \right]}{\sigma_{l} \left( l_t\right)}  > 0 \right] \epsilon^{X}_t + \Gamma' Z_{i,t-1}  + e_{i,t}$. The second row results come from estimating the specification $\log k_{i,t} - \log k_{i,t-1} = \alpha_i + \alpha_{s,t} + \beta \mathbbm{1} \left[\frac{l_{i,t} - \mathbb{E}_i \left[l_t \right]}{\sigma_{l} \left( l_t\right)}  > 1 \right] \epsilon^{X}_t + \Gamma' Z_{i,t-1}  + e_{i,t}$. Standard errors are clustered at the firm and time level.}
\end{table}
Each row of the table presents the results for a different specification of the indicator functions. Across the different specifications we obtain mix results. For both specifications of the indicator functions results are negative and statistically different from zero for the first two specifications. Furthermore, the more stringent specification of the indicator function with leverage one standard deviation above its standardized mean, the estimated coefficients are almost twice as large compared to the less stringent specification. However, the latter two specifications which introduce a greater amount of controls do not find estimates significantly different from zero.  

Table \ref{tab:appendix_indicator_average} presents the results of estimating Equation \ref{eq:baseline_heterogeneous_average} under the two indicator function specifications. 
\begin{table}[ht]
    \centering
    \caption{Heterogeneous Response of Investment to $\epsilon^{X}_t$ \\ \footnotesize Alt. Specification - Ind. Functions}
    \label{tab:appendix_indicator_average}
    \scriptsize
\begin{tabular}{lcccc}
 & \multicolumn{4}{c}{Firm investment - $\Delta \log k_{i,t}$ } \\
  & (1) & (2) & (3) & (4) \\ \hline \hline
 &  &  &  &  \\
$\mathbbm{1} \left[\frac{l_{i,t} - \mathbb{E}_i \left[l_t \right]}{\sigma_{l} \left( l_t\right)}  > 0 \right]$ & -0.375*** & -0.362*** & -0.266 & -0.213 \\
 & (0.120) & (0.129) & (0.174) & (0.175) \\
 &  &  &  &  \\
$\mathbbm{1} \left[\frac{l_{i,t} - \mathbb{E}_i \left[l_t \right]}{\sigma_{l} \left( l_t\right)}  > 1 \right]$ & -0.636* & -0.660* & -0.486 & -0.429 \\
 & (0.350) & (0.343) & (0.338) & (0.341) \\
 &  &  &  &  \\
Firm FE & yes & yes & yes & yes \\
Sector-Time FE & yes & yes & yes & yes \\
Control for Lagged Inv. & no & yes & yes & yes \\
Firm-Controls & no & no & yes & yes \\
Aggregate-Controls & no & no & no & yes \\
 &  &  &  &  \\
Observations & 12,646 & 12,473 & 11,869 & 11,863 \\ \hline \hline
\multicolumn{5}{c}{ Clustered standard errors in parentheses} \\
\multicolumn{5}{c}{ *** p$<$0.01, ** p$<$0.05, * p$<$0.1} \\
\end{tabular}
\floatfoot{\textbf{Note:}  Results in each panel represent results of different regression. The results presented are estimates of parameter $\beta$ of one of the following two specifications. The top panel presents results from estimating $\Delta \log k_{i,t} = \alpha_i + \alpha_{s,q} + \gamma \epsilon^{X}_t + \beta \mathbbm{1} \left[\frac{l_{i,t} - \mathbb{E}_i \left[l_t \right]}{\sigma_{l} \left( l_t\right)}  > 0 \right] \epsilon^{X}_t + \Gamma'_1 Z_{i,t-1} + \Gamma'_2 Y_{t-1}  + \tilde{e}_{i,t}$. The bottom panel presents results from estimating $\Delta \log k_{i,t} = \alpha_i + \alpha_{s,q} + \gamma \epsilon^{X}_t + \beta \mathbbm{1} \left[\frac{l_{i,t} - \mathbb{E}_i \left[l_t \right]}{\sigma_{l} \left( l_t\right)}  > 1 \right] \epsilon^{X}_t + \Gamma'_1 Z_{i,t-1} + \Gamma'_2 Y_{t-1}  + \tilde{e}_{i,t}$. Standard errors are clustered at the firm and time level.  }
\end{table}
Results are in line with the results presented in Table \ref{tab:appendix_indicator}. Results are negative and statistically significantly different from zero for the first two specifications for both indicator function specifications. Again, the results are greater in magnitude for the more stringent indicator function specification than for the less stringent indicator function specification. At the same time, results become non-significant for the two last specifications which include a greater amount of controls. 

Overall, we interpret the mix results presented in Tables \ref{tab:appendix_indicator} and \ref{tab:appendix_indicator_average} positively. While results are not significant across all of the specifications, estimated coefficients are negative and thus in line with the results presented in Section \ref{sec:firm_evidence}. Furthermore, the indicator function specifications which takes the value of one when the standardized measure of leverage is one standard deviation above its mean yields estimates which are twice as big compared to the indicator function specification which takes the value of one when the standardized measure of leverage is positive, i.e. leverage above its sample mean.

%%%%%%%%%%%%%%%%%%%
% Definicion alternativa de leverage

Next, we carry out two additional robustness check using alternative measures of leverage. In particular, we consider three alternative measures of leverage: (i) ratio of short-term or under a year debt to total assets, (ii) ratio of long-term or above a year debt to total assets, (iii) ratio of debt with banks and/or other domestic financial institutions to total assets.
\begin{table}[ht]
    \centering
    \caption{Heterogeneous Response of Investment  to $\epsilon^{X}_t$ \\ \footnotesize Short term Leverage }
    \scriptsize
    \label{tab:baseline_heterogeneous_def2}
\begin{tabular}{lcccc}
 & \multicolumn{4}{c}{Firm investment - $\Delta \log k_{i,t}$ } \\
  & (1) & (2) & (3) & (4) \\ \hline \hline
 &  &  &  &  \\
$\left(\tilde{l}_{i,t-1} - \mathbb{E}_{i} \left[ \tilde{l}_{t}\right] \right) \epsilon^{X}_t$  & -0.222* & -0.225* & -0.159 & -0.165 \\
 & (0.112) & (0.113) & (0.142) & (0.141) \\
 &  &  &  &  \\
Firm FE & yes & yes & yes & yes \\
Sector-Time FE & yes & yes & yes & yes \\
Control for Lagged Inv. & no & yes & yes & yes \\
Control for Lagged Currency Mismatch & no & no & yes & yes \\
Control for $l_{i,t-1}$ & no & no & no & yes \\
 &  &  &  &  \\
Observations & 12,617 & 12,448 & 11,936 & 11,842 \\ \hline 
$R^{2}$ & 0.142 & 0.152 & 0.149 & 0.162 \\ \hline \hline
\multicolumn{5}{c}{ Clustered standard errors in parentheses} \\
\multicolumn{5}{c}{ *** p$<$0.01, ** p$<$0.05, * p$<$0.1} \\
\end{tabular}
\floatfoot{\textbf{Note:} Results from estimating $\log k_{i,t} - \log k_{i,t-1} = \alpha_i + \alpha_{s,t} + \beta \left(\tilde{l}_{i,t-1} - \mathbb{E}_{i} \left[ \tilde{l}_{t}\right] \right) \epsilon^{X}_t + \Gamma' Z_{i,t-1}  + e_{i,t}$, where $\tilde{l}$ reflects our alternative definition of leverage, i.e., current liabilities to total assets. We have standardized $\left(\tilde{l}_{i,t-1} - \mathbb{E}_{i} \left[ \tilde{l}_{t}\right] \right)$ over the entire sample using firm specific standard deviations. Standard errors are clustered at the firm and time level. }
\end{table}
These different leverage specifications are motivated by the differences in borrowing patterns between Emerging Market and US companies. While US companies tend to primarily borrow long term, through commercial papers and not through banks, Emerging Market economies' companies tend to borrow at lower maturities, relying relatively more in domestic banks to fund working capital and capital expenditures.

First, we test whether the heterogeneous impact of US interest rate shocks, found in our benchmark specification in Table \ref{tab:baseline_heterogeneous}, is driven by firms' short-term leverage. 
\begin{table}[ht]
    \centering
    \caption{Heterogeneous Response of Investment to $\epsilon^{X}_t$ \\ \footnotesize S.T. Leverage \& Alt. Specification}
    \label{tab:baseline_heterogeneous_average_def2}
    \scriptsize
\begin{tabular}{lcccc}
 & \multicolumn{4}{c}{Firm investment - $\Delta \log k_{i,t}$ } \\
  & (1) & (2) & (3) & (4) \\ \hline \hline
 &  &  &  &  \\
$\left(l_{i,t-1} - \mathbb{E}_{i} \left[ l_{t}\right] \right) \epsilon^{X}_t$ & -0.207* & -0.209* & -0.160 & -0.156 \\
 & (0.106) & (0.107) & (0.131) & (0.130) \\
 &  &  &  &  \\
Firm FE & yes & yes & yes & yes \\
Sector-Time FE & yes & yes & yes & yes \\
Control for Lagged Inv. & no & yes & yes & yes \\
Firm-Controls & no & no & yes & yes \\
Aggregate-Controls & no & no & no & yes \\
 &  &  &  &  \\
Observations & 12,617 & 12,448 & 11,849 & 11,843 \\ \hline
$R^{2}$ & 0.056 & 0.064 & 0.075 & 0.081 \\ \hline \hline
\multicolumn{5}{c}{ Clustered standard errors in parentheses} \\
\multicolumn{5}{c}{ *** p$<$0.01, ** p$<$0.05, * p$<$0.1} \\
\end{tabular}
\floatfoot{\textbf{Note:} Results from estimating $\Delta \log k_{i,t} = \alpha_i + \alpha_{s,q} + \gamma \epsilon^{X}_t + \beta \left(l_{i,t-1} - \mathbb{E}_{i} \left[ l_{t}\right] \right) \epsilon^{X}_t + \Gamma'_1 Z_{i,t-1} + \Gamma'_2 Y_{t-1}  + \tilde{e}_{i,t}$. We have standardized $\left(l_{i,t-1} - \mathbb{E}_{i} \left[ l_{t}\right] \right)$ over the entire sample using firm specific standard deviations. Standard errors are clustered at the firm and time level.  }
\end{table}
Results for both the benchmark and alternative econometric specifications are presented in Tables \ref{tab:baseline_heterogeneous_def2} and \ref{tab:baseline_heterogeneous_average_def2}. These tables lead to two conclusions. First, the estimated heterogeneous impact is quantitatively smaller than those estimated in the benchmark specifications in Tables \ref{tab:baseline_heterogeneous} and \ref{tab:baseline_heterogeneous_average}. This may be driven by short-term leverage being less informative for a firms' financial status (closeness to a borrowing constraint or closeness to default). Second, results are not statistically different from zero for all empirical specifications. In particular, for both econometric specifications, results are different from zero for specifications with a small number of controls, but lose their statistical significance for the last two specifications.

Second, we test the role of long term leverage in explaining the heterogeneous response of firms' investment to US interest rate shocks. 
\begin{table}[ht]
    \centering
    \caption{Heterogeneous Response of Investment  to $\epsilon^{X}_t$ \\ \footnotesize Long term Leverage }
    \scriptsize
    \label{tab:baseline_heterogeneous_def3}
\begin{tabular}{lcccc}
 & \multicolumn{4}{c}{Firm investment - $\Delta \log k_{i,t}$ } \\
  & (1) & (2) & (3) & (4) \\ \hline \hline
 &  &  &  &  \\
$\left(\tilde{l}_{i,t-1} - \mathbb{E}_{i} \left[ \tilde{l}_{t}\right] \right) \epsilon^{X}_t$   & -0.278*** & -0.275*** & -0.205* & -0.202* \\
 & (0.0961) & (0.0936) & (0.114) & (0.112) \\
 &  &  &  &  \\
Firm FE & yes & yes & yes & yes \\
Sector-Time FE & yes & yes & yes & yes \\
Control for Lagged Inv. & no & yes & yes & yes \\
Control for Lagged Currency Mismatch & no & no & yes & yes \\
Control for $l_{i,t-1}$ & no & no & no & yes \\
 &  &  &  &  \\
Observations & 11,670 & 11,516 & 10,064 & 9,995 \\ \hline 
$R^{2}$ & 0.149 & 0.155 & 0.161 & 0.170 \\ \hline \hline
\multicolumn{5}{c}{ Clustered standard errors in parentheses} \\
\multicolumn{5}{c}{ *** p$<$0.01, ** p$<$0.05, * p$<$0.1} \\
\end{tabular}
\floatfoot{\textbf{Note:} Results from estimating $\log k_{i,t} - \log k_{i,t-1} = \alpha_i + \alpha_{s,t} + \beta \left(\tilde{l}_{i,t-1} - \mathbb{E}_{i} \left[ \tilde{l}_{t}\right] \right) \epsilon^{X}_t + \Gamma' Z_{i,t-1}  + e_{i,t}$, where $\tilde{l}$ reflects our alternative definition of leverage, i.e., current liabilities to total assets. We have standardized $\left(\tilde{l}_{i,t-1} - \mathbb{E}_{i} \left[ \tilde{l}_{t}\right] \right)$ over the entire sample using firm specific standard deviations. Standard errors are clustered at the firm and time level. }
\end{table}
\begin{table}[ht]
    \centering
    \caption{Heterogeneous Response of Investment to $\epsilon^{X}_t$ \\ \footnotesize L.T. Leverage \& Alt. Specification}
    \label{tab:baseline_heterogeneous_average_def3}
    \scriptsize
\begin{tabular}{lcccc}
 & \multicolumn{4}{c}{Firm investment - $\Delta \log k_{i,t}$ } \\
  & (1) & (2) & (3) & (4) \\ \hline \hline
 &  &  &  &  \\
$\left(l_{i,t-1} - \mathbb{E}_{i} \left[ l_{t}\right] \right) \epsilon^{X}_t$ & -0.270*** & -0.270*** & -0.240** & -0.229** \\
 & (0.0977) & (0.0954) & (0.0981) & (0.0957) \\
 &  &  &  &  \\
Firm FE & yes & yes & yes & yes \\
Sector-Time FE & yes & yes & yes & yes \\
Control for Lagged Inv. & no & yes & yes & yes \\
Firm-Controls & no & no & yes & yes \\
Aggregate-Controls & no & no & no & yes \\
 &  &  &  &  \\
Observations & 11,670 & 11,516 & 10,919 & 10,913 \\ \hline
$R^{2}$ & 0.044 & 0.046 & 0.051 & 0.061 \\ \hline \hline
\multicolumn{5}{c}{ Clustered standard errors in parentheses} \\
\multicolumn{5}{c}{ *** p$<$0.01, ** p$<$0.05, * p$<$0.1} \\
\end{tabular}
\floatfoot{\textbf{Note:} Results from estimating $\Delta \log k_{i,t} = \alpha_i + \alpha_{s,q} + \gamma \epsilon^{X}_t + \beta \left(l_{i,t-1} - \mathbb{E}_{i} \left[ l_{t}\right] \right) \epsilon^{X}_t + \Gamma'_1 Z_{i,t-1} + \Gamma'_2 Y_{t-1}  + \tilde{e}_{i,t}$. We have standardized $\left(l_{i,t-1} - \mathbb{E}_{i} \left[ l_{t}\right] \right)$ over the entire sample using firm specific standard deviations. Standard errors are clustered at the firm and time level.  }
\end{table}
Tables \ref{tab:baseline_heterogeneous_def3} and \ref{tab:baseline_heterogeneous_average_def3} present the results for both main econometric specifications presented in Equations \ref{eq:baseline_heterogeneous} and \ref{eq:baseline_heterogeneous_average}. Alike the results presented for short-term leverage, the estimated heterogeneity is smaller in magnitude than that estimated in Section \ref{sec:firm_evidence}. Unlike the results presented for short-term leverage, results are statistically different from zero for all empirical specifications considered. Consequently, these results suggest that long-term leverage plays a key role in explaining the transmission of US interest rate shocks into firms' investment. 

Finally, we test the role of bank debt leverage in explaining the heterogeneity in the impact of US interest rate shocks in firms' investment. 
\begin{table}[ht]
    \centering
    \caption{Heterogeneous Response of Investment  to $\epsilon^{X}_t$ \\ \footnotesize Bank Leverage }
    \scriptsize
    \label{tab:baseline_heterogeneous_def4}
\begin{tabular}{lcccc}
 & \multicolumn{4}{c}{Firm investment - $\Delta \log k_{i,t}$ } \\
  & (1) & (2) & (3) & (4) \\ \hline \hline
 &  &  &  &  \\
$\left(\tilde{l}_{i,t-1} - \mathbb{E}_{i} \left[ \tilde{l}_{t}\right] \right) \epsilon^{X}_t$  & -0.126 & -0.133 & -0.157* & -0.159* \\
 & (0.0826) & (0.0843) & (0.0805) & (0.0854) \\
 &  &  &  &  \\
Firm FE & yes & yes & yes & yes \\
Sector-Time FE & yes & yes & yes & yes \\
Control for Lagged Inv. & no & yes & yes & yes \\
Control for Lagged Currency Mismatch & no & no & yes & yes \\
Control for $l_{i,t-1}$ & no & no & no & yes \\
 &  &  &  &  \\
Observations & 11,022 & 10,917 & 10,405 & 10,329 \\ \hline 
$R^{2}$ & 0.148 & 0.153 & 0.155 & 0.165 \\ \hline \hline
\multicolumn{5}{c}{ Clustered standard errors in parentheses} \\
\multicolumn{5}{c}{ *** p$<$0.01, ** p$<$0.05, * p$<$0.1} \\
\end{tabular}
\floatfoot{\textbf{Note:} Results from estimating $\log k_{i,t} - \log k_{i,t-1} = \alpha_i + \alpha_{s,t} + \beta \left(\tilde{l}_{i,t-1} - \mathbb{E}_{i} \left[ \tilde{l}_{t}\right] \right) \epsilon^{X}_t + \Gamma' Z_{i,t-1}  + e_{i,t}$, where $\tilde{l}$ reflects our alternative definition of leverage, i.e., current liabilities to total assets. We have standardized $\left(\tilde{l}_{i,t-1} - \mathbb{E}_{i} \left[ \tilde{l}_{t}\right] \right)$ over the entire sample using firm specific standard deviations. Standard errors are clustered at the firm and time level. }
\end{table}
\begin{table}[ht]
    \centering
    \caption{Heterogeneous Response of Investment to $\epsilon^{X}_t$ \\ \footnotesize Bank Leverage \& Alt. Specification}
    \label{tab:baseline_heterogeneous_average_def4}
    \scriptsize
\begin{tabular}{lcccc}
 & \multicolumn{4}{c}{Firm investment - $\Delta \log k_{i,t}$ } \\
  & (1) & (2) & (3) & (4) \\ \hline \hline
 &  &  &  &  \\
$\left(l_{i,t-1} - \mathbb{E}_{i} \left[ l_{t}\right] \right) \epsilon^{X}_t$ & -0.0959 & -0.106 & -0.0958 & -0.0619 \\
 & (0.0728) & (0.0753) & (0.0859) & (0.0874) \\
 &  &  &  &  \\
Firm FE & yes & yes & yes & yes \\
Sector-Time FE & yes & yes & yes & yes \\
Control for Lagged Inv. & no & yes & yes & yes \\
Firm-Controls & no & no & yes & yes \\
Aggregate-Controls & no & no & no & yes \\
 &  &  &  &  \\
Observations & 11,042 & 10,937 & 10,349 & 10,346 \\ \hline
$R^{2}$ & 0.043 & 0.046 & 0.055 & 0.064 \\ \hline \hline
\multicolumn{5}{c}{ Clustered standard errors in parentheses} \\
\multicolumn{5}{c}{ *** p$<$0.01, ** p$<$0.05, * p$<$0.1} \\
\end{tabular}
\floatfoot{\textbf{Note:} Results from estimating $\Delta \log k_{i,t} = \alpha_i + \alpha_{s,q} + \gamma \epsilon^{X}_t + \beta \left(l_{i,t-1} - \mathbb{E}_{i} \left[ l_{t}\right] \right) \epsilon^{X}_t + \Gamma'_1 Z_{i,t-1} + \Gamma'_2 Y_{t-1}  + \tilde{e}_{i,t}$. We have standardized $\left(l_{i,t-1} - \mathbb{E}_{i} \left[ l_{t}\right] \right)$ over the entire sample using firm specific standard deviations. Standard errors are clustered at the firm and time level.  }
\end{table}
Tables \ref{tab:baseline_heterogeneous_def4} and \ref{tab:baseline_heterogeneous_average_def4} present the results for both main econometric specifications presented in Equations \ref{eq:baseline_heterogeneous} and \ref{eq:baseline_heterogeneous_average}. Across the two tables, the results are both smaller in magnitude and mostly not significantly different from zero (the exemption being the last two specifications of Table \ref{tab:baseline_heterogeneous_def4}). These results may suggest that firms which primarily borrow from banks and not from capital markets are less exposed to US interest rate shocks. %These result is also in line with the additional SVAR results presented in Section \ref{sec:supporting_evidence}, which showed that inter-bank rates do not increase in response to US monetary policy shocks, while market rates, particularly those for borrowing in foreign currency, exhibit a sharp increase. 

Lastly, we show that the firm-level results presented in Tables \ref{tab:baseline_heterogeneous} and \ref{tab:baseline_heterogeneous_average} are robust to a different identification scheme. In particular, we test whether our results are robust to using a similar approach to the one used by \cite{ottonello2020financial}, which relies only on the high-frequency movement of the Fed Fund Futures around FOMC meetings. We denote this high-frequency surprise by $\epsilon^{FFF}_t$ constructed as
\begin{align} \label{eq:shock_construction_appendix}
    \epsilon^{FFF}_t = FFF_{t+\Delta_{+}}-FFF_{t+\Delta_{-}}
\end{align}
where $t$ is the time of the monetary announcement, $FFF_t$ is the Federal Funds Futures contract for the current month at time $t$, $\Delta_{+}$ and $\Delta_{-}$ control the size of the time window around the announcement. We follow \cite{nakamura2018high} and use a time window of 30 minutes, from 10 minutes before the FOMC announcement ($\Delta_{-}=10$ minutes) to 20 minutes after it ($\Delta_{+}=20$ minutes). This high-frequency shocks are sourced from \cite{acosta2020estimating}. We average the shocks within a same quarter. Tab

\begin{table}[ht]
    \centering
    \caption{Heterogeneous Response of Investment  to $\epsilon^{X}_t$}
    \footnotesize
    \label{tab:baseline_heterogeneous_FFF}
\begin{tabular}{lcccc}
 & \multicolumn{4}{c}{Firm investment - $\Delta \log k_{i,t}$ } \\
  & (1) & (2) & (3) & (4) \\ \hline \hline
 &  &  &  &  \\
$\left(l_{i,t-1} - \mathbb{E}_{i} \left[ l_{t}\right] \right) \epsilon^{X}_t$ & -0.195** & -0.219*** & -0.301*** & -0.334*** \\
 & (0.0775) & (0.0778) & (0.0813) & (0.0815) \\
 &  &  &  &  \\
Firm FE & yes & yes & yes & yes \\
Sector-Time FE & yes & yes & yes & yes \\
Control for Lagged Inv. & no & yes & yes & yes \\
Control for Lagged Currency Mismatch & no & no & yes & yes \\
Control for $l_{i,t-1}$ & no & no & no & yes \\
 &  &  &  &  \\
Observations & 12,183 & 12,011 & 11,508 & 11,411 \\  \hline
 $R^{2}$ & 0.141 & 0.152 & 0.150 & 0.164 \\ \hline \hline
\multicolumn{5}{c}{ Clustered standard errors in parentheses} \\
\multicolumn{5}{c}{ *** p$<$0.01, ** p$<$0.05, * p$<$0.1} \\
\end{tabular}
\floatfoot{\textbf{Note:} Results from estimating $\log k_{i,t} - \log k_{i,t-1} = \alpha_i + \alpha_{s,t} + \beta \left(l_{i,t-1} - \mathbb{E}_{i} \left[ l_{t}\right] \right) \epsilon^{X}_t + \Gamma' Z_{i,t-1}  + e_{i,t}$. We have standardized $\left(l_{i,t-1} - \mathbb{E}_{i} \left[ l_{t}\right] \right)$ over the entire sample using firm specific standard deviations. Standard errors are clustered at the firm-level.  }
\end{table}
\begin{table}[ht]
    \centering
    \caption{Heterogeneous Response of Investment to $\epsilon^{X}_t$ \\ \footnotesize Average \& Differential Impact}
    \label{tab:baseline_heterogeneous_average_FFF}
    \footnotesize
\begin{tabular}{lcccc}
 & \multicolumn{4}{c}{Firm investment - $\Delta \log k_{i,t}$ } \\
  & (1) & (2) & (3) & (4) \\ \hline \hline
 &  &  &  &  \\
$\left(l_{i,t-1} - \mathbb{E}_{i} \left[ l_{t}\right] \right) \epsilon^{X}_t$ & -0.134* & -0.160** & -0.306*** & -0.298*** \\
 & (0.0752) & (0.0758) & (0.0798) & (0.0806) \\
 &  &  &  &  \\
Firm FE & yes & yes & yes & yes \\
Sector-Time FE & yes & yes & yes & yes \\
Control for Lagged Inv. & no & yes & yes & yes \\
Firm-Controls & no & no & yes & yes \\
Aggregate-Controls & no & no & no & yes \\
 &  &  &  &  \\
Observations & 12,183 & 12,011 & 11,411 & 11,123 \\ \hline
$R^{2}$ & 0.056 & 0.064 & 0.077 & 0.080 \\ \hline \hline
\multicolumn{5}{c}{ Clustered standard errors in parentheses} \\
\multicolumn{5}{c}{ *** p$<$0.01, ** p$<$0.05, * p$<$0.1} \\
\end{tabular}
\floatfoot{\textbf{Note:} Results from estimating $\Delta \log k_{i,t} = \alpha_i + \alpha_{s,q} + \gamma \epsilon^{X}_t + \beta \left(l_{i,t-1} - \mathbb{E}_{i} \left[ l_{t}\right] \right) \epsilon^{X}_t + \Gamma'_1 Z_{i,t-1} + \Gamma'_2 Y_{t-1}  + \tilde{e}_{i,t}$. We have standardized $\left(l_{i,t-1} - \mathbb{E}_{i} \left[ l_{t}\right] \right)$ over the entire sample using firm specific standard deviations. Standard errors are clustered at the firm-level.  }
\end{table}

%%%%%%%%%%%%%%%%%%%%%%%%%%%%%%%%%%%%%%%%%%%%%%%%%%%%%%%%%%%%%%%%%%%%
\newpage
\section{Additional Details: Two Period Structural Model } \label{sec:appendix_additional_structural_model_details}

We can state entrepreneur $i$'s utility maximization problem as 
\begin{equation*}
    \begin{aligned}
    \max_{c_{i,0},c_{i,1},k_{i,1},b_{i,1}} \quad &  \mathcal{U} = \ln c_{i,0} + \beta \ln c_{i,1} \\
    \textrm{s.t.} \quad & c_{i,0} + k_{i,1} = k^{\alpha}_{i,0} + b_{i,1} - b_{i,0} \left(1+r_0\right) \\
    & c_{i,1} = k^{\alpha}_{i,1} - b_{i,1} \left(1+r_1\right) \\
      & b_{i,1} \leq \theta_i k_{i,1}
    \end{aligned}
\end{equation*}
The entrepreneur utility maximization problem's solution can be characterized by the Euler equations for capital $k_{i,1}$ and debt $b_{i,1}$
\begin{align*}
    \frac{1}{c_{i,0}} &= \frac{\beta}{c_{i,1}} \alpha k^{\alpha-1}_{i,1} + \mu \theta  \\
    \frac{1}{c_{i,0}} &= \frac{\beta}{c_{i,1}} \left(1+r_1\right) + \mu
\end{align*}
where $\mu$ represents the Lagrange multiplier on the leverage constraint in Equation \ref{eq:leverage_constraint}. The solution to the constrained entrepreneur's problem is characterized by the two Euler equations, the two budget constraints and the complementary slackness condition. Exploiting the fact that the budget constraint binds, i.e. $b_1 = \theta k_1$, we can use the budget constraints to express $c_0$ and $c_1$ as a function of $k_1$
\begin{align*}
    c_0 &= k^{\alpha}_0 - b_0 \left(1+r_0\right) - k_1 \left(1-\theta \right) \\
    c_1 &= k^{\alpha}_1 - b_1 \left(1+r_1\right)
\end{align*}
Then, we can use the Euler Equation for bonds to obtain an expression for $\mu$ that depends only on $k_1$
\begin{align*}
    \frac{1}{c_{i,0}} &= \frac{\beta}{c_{i,1}} \left(1+r_1\right) + \mu \\
    \mu &= \frac{1}{c_{i,0}} - \frac{\beta}{c_{i,1}} \left(1+r_1\right) \\
    \mu &= \frac{1}{k^{\alpha}_0 - b_0 \left(1+r_0\right) - k_1 \left(1-\theta \right)} - \frac{\beta \left(1+r_1\right)}{k^{\alpha}_1 - b_1 \left(1+r_1\right)}
\end{align*}
Finally, we can use the expression for optimal capital accumulation for the constrained entrepreneur in Equation \ref{eq:optimal_capital_constraint}, to obtain an expression for capital depending only on parameters
\begin{align}
    & \alpha k^{\alpha-1}_1 - \left(1+r_1\right) - \mu \frac{c_1}{\beta} \left(1-\theta\right) = 0 \nonumber \\
    & \alpha k^{\alpha-1}_1 - \left(1+r_1\right) - \left[\frac{1}{k^{\alpha}_0 - b_0 \left(1+r_0\right) - k_1 \left(1-\theta \right)} - \frac{\beta \left(1+r_1\right)}{k^{\alpha}_1 - \theta k_1 \left(1+r_1\right)} \right] \frac{c_1}{\beta} \left(1-\theta\right) = 0 \nonumber \\
    & \alpha k^{\alpha-1}_1 - \left(1+r_1\right) - \left[\frac{1}{k^{\alpha}_0 - b_0 \left(1+r_0\right) - k_1 \left(1-\theta \right)} - \frac{\beta \left(1+r_1\right)}{k^{\alpha}_1 - \theta k_1 \left(1+r_1\right)} \right] \frac{1}{\beta} \left(1-\theta\right) \{k^{\alpha}_1 - b_1 \left(1+r_1\right) \} = 0 \label{eq:implicit_function_constrained}
\end{align}

\noindent
\textbf{Proof of Proposition 1:} Proposition 1 states that for a given set of parameters $\{\alpha,\theta\}$, initial capital stock $k_0$, and close enough initial stock of debt $b_0$ such that one entrepreneur is unconstrained and a second one is constrained, the response of the optimal stock of capital $k_1$ with respect to the gross interest rate $r_1$ is greater in magnitude for the unconstrained entrepreneur than for the constrained entrepreneur.

We can use the expression in Equation \ref{eq:implicit_function_constrained} combined with the implicit function theorem to characterize the response of capital accumulation to changes in the interest rate $(1+r_1)$ and to the tightness of the capital accumulation $\theta$. Denoting function $F\left(k_1,1+r_1,\theta\right)$ as \small
\begin{align}
   F\left(k_1,1+r_1,\theta\right) &=  \alpha k^{\alpha-1}_1 - \left(1+r_1\right) - \nonumber \\
   & \left[\frac{1}{k^{\alpha}_0 - b_0 \left(1+r_0\right) - k_1 \left(1-\theta \right)} - \frac{\beta \left(1+r_1\right)}{k^{\alpha}_1 - \theta k_1 \left(1+r_1\right)} \right] \frac{1}{\beta} \left(1-\theta\right) \{k^{\alpha}_1 - \theta k_1 \left(1+r_1\right) \} = 0
\end{align} \normalsize
we can compute the derivatives of interest as
\begin{align*}
    \frac{\partial k_1}{\partial 1+r_1} &= - \frac{F'_{1+r_1}}{F'_{k_1}} \\
    %\frac{\partial k_1}{\partial \theta} &= - \frac{F'_{\theta}}{F'_{k_1}}
\end{align*}

We start by computing the derivatives of interest for the case of the unconstrained entrepreneur. In this case, the third term of Equation \ref{eq:implicit_function_constrained} is zero as the Lagrange multiplier on the leverage constraint is zero. Thus, the function which describes the optimal capital accumulation becomes
\begin{align*}
    F\left(k_1,1+r_1,\theta\right) &=  \alpha k^{\alpha-1}_1 - \left(1+r_1\right) = 0
\end{align*}
We can compute the derivative of the optimal capital accumulation to the change in interest rate $\left(1+r_1\right)$ as 
\begin{align*}
    \frac{\partial k_1}{\partial 1+r_1} &= - \frac{F'_{1+r_1}}{F'_{k_1}}
\end{align*}
using once again the implicit function theorem. Then,
\begin{align}
    \frac{\partial k_1}{\partial 1+r_1} &= - \frac{F'_{1+r_1}}{F'_{k_1}} = - \frac{-1}{\alpha \times \left(\alpha-1\right) k^{\alpha-2}_1} = \frac{1}{\alpha \times \left(\alpha-1\right) k^{\alpha-2}_1} 
\end{align}
First, note that the denominator or $\alpha \times \left(\alpha-1\right) k^{\alpha-2}_1$ is negative as $\alpha-1<0$. Thus, the unconstrained optimal capital accumulation decreases as the interest rate $(1+r_1)$ increases. 

Next, we turn to computing the response of the constrained entrepreneur's capital accumulation $k_1$ to changes in the interest rate $1+r_1$. We start by computing the derivative of function $F\left(k_1,1+r_1,\theta\right)$ with respect to $1+r_1$. The terms that depend on $(1+r_1)$ in function $F$ are expressed below
\begin{align*}
    - \left(1+r_1\right) - \left[\frac{1}{k^{\alpha}_0 - b_0 \left(1+r_0\right) - k_1 \left(1-\theta \right)} - \frac{\beta \left(1+r_1\right)}{k^{\alpha}_1 - \theta k_1 \left(1+r_1\right)} \right] \frac{1}{\beta} \left(1-\theta\right) \{k^{\alpha}_1 - \theta k_1 \left(1+r_1\right) \}
\end{align*}
Note that the first term, $-\left(1+r_1\right)$ is the same for both the unconstrained and the constrained cases. Thus, any differences will arise from the derivative of the second term with respect to $\left(1+r_1\right)$:
\begin{align*}
    - \frac{\left(1-\theta\right)}{\beta} \times \left[\frac{1}{k^{\alpha}_0 - b_0 \left(1+r_0\right) - k_1 \left(1-\theta \right)} - \frac{\beta \left(1+r_1\right)} {k^{\alpha}_1 - \theta k_1 \left(1+r_1\right)} \right]   \times \{k^{\alpha}_1 - \theta k_1 \left(1+r_1\right) \}
\end{align*}
Note that the second and third factors in the term above depend on $\left(1+r_1\right)$. The derivative of this term of $F\left(k_1,{1+r_1},\theta\right)$ is equal to
\begin{align*}
    & \left(-1\right) \frac{1-\theta}{\beta} \bigg\{ \{k^{\alpha}_1 - \theta k_1 \left(1+r_1\right) \} \times \left(-\beta\right) \times \left[\frac{k^{\alpha}_1 - \theta k_1\left(1+r_1\right)-\left(1+r_1\right) \times \left(-\theta k_1\right)}{\{k^{\alpha}_1 - \theta k_1 \left(1+r_1\right)\}^2} \right] \\
    & \qquad +   \left[\frac{1}{k^{\alpha}_0 - b_0 \left(1+r_0\right) - k_1 \left(1-\theta \right)} - \frac{\beta \left(1+r_1\right)} {k^{\alpha}_1 - \theta k_1 \left(1+r_1\right)} \right]   \times \left( -\theta k_1 \right)  \bigg\} 
\end{align*}
To determine the sign of this expression we individually consider every term inside the big curly brackets. The first term is negative
\begin{align*}
    & \{k^{\alpha}_1 - \theta k_1 \left(1+r_1\right) \} \times \left(-\beta\right) \times \left[\frac{k^{\alpha}_1 - \theta k_1\left(1+r_1\right)-\left(1+r_1\right) \times \left(-\theta k_1\right)}{\{k^{\alpha}_1 - \theta k_1 \left(1+r_1\right)\}^2} \right] \\
    & \underbrace{\{k^{\alpha}_1 - \theta k_1 \left(1+r_1\right) \}}_{>0} \times \underbrace{\left(-\beta\right)}_{<0} \times \underbrace{\left[\frac{k^{\alpha}_1}{\{k^{\alpha}_1 - \theta k_1 \left(1+r_1\right)\}^2} \right]}_{>0} < 0
\end{align*}
The second term is also negative
\begin{align*}
    \underbrace{\left[\frac{1}{k^{\alpha}_0 - b_0 \left(1+r_0\right) - k_1 \left(1-\theta \right)} - \frac{\beta \left(1+r_1\right)} {k^{\alpha}_1 - \theta k_1 \left(1+r_1\right)} \right]}_{>0}   \times \underbrace{\left( -\theta k_1 \right)}_{<0} < 0
\end{align*}
Thus, the whole term within the curly brackets is negative, implying that the derivative with respect to the gross interest rate $\left(1+r_1\right)$ of the last term of constrained entrepreneurs in Equation \ref{eq:implicit_function_constrained}. Consequently,
\begin{align*}
    |\left(F'_{1+r_1}\right)^{\text{Unconstrained}}| > |\left(F'_{1+r_1}\right)^{\text{Constrained}}| 
\end{align*}
as long as the positive term of the derivative for the constrained entrepreneur is not large enough to switch the sign of the whole derivative. In that case, capital accumulation would become increasing in the gross interest rate, a case which is not of economic interest, and which still imply that capital accumulation of unconstrained entrepreneurs would decrease by more than that of constrained entrepreneurs.

In order to compute the derivative of function $F\left(k_1,1+r_1,\theta\right)$ with respect to $k_1$ we separately compute the derivative of the different terms of function $F$. First,
\begin{align*}
    \frac{\partial \left[k^{\alpha}_0 - b_0 \left(1+r_0\right) - k_1 \left(1-\theta \right)\right]^{-1} }{\partial k_1} &= \left(1-\theta\right) \left[\underbrace{k^{\alpha}_0 - b_0 \left(1+r_0\right) - k_1 \left(1-\theta \right)}_{c_0}\right]^{-2}
\end{align*}
which is strictly positive as $(1-\theta)>0$ given our assumptions over $\theta$ and given the logarithmic preferences, $c_0>0$. Next,
\begin{align*}
    \frac{\partial \beta \left(1+r_1\right) \left[ k^{\alpha}_1 - \theta k_1 \left(1+r_1\right) \right]^{-1} }{\partial k_1} &= - \beta \left(1+r_1\right) \left[ \underbrace{k^{\alpha}_1 - \theta k_1 \left(1+r_1\right)}_{c_1} \right]^{-2} \times \left[ \alpha k^{\alpha-1} - \theta \left(1+r_1\right) \right]
\end{align*}
which is also strictly negative given that $\beta \left(1+r_1\right)>0$, $c_1>0$ given the logarithmic preferences, and $\alpha k^{\alpha-1} - \theta \left(1+r_1\right) >0$ given that $\alpha k^{\alpha-1} - \left(1+r_1\right)>0$ as the entrepreneur is constrained and $\theta \in (0,1)$. Lastly, the derivative of term $\{k^{\alpha}_1 - \theta_1 \left(1+r_1\right) \}$
is
\begin{align*}
    \frac{\partial k^{\alpha}_1 - \theta k_1 \left(1+r_1\right)}{\partial k_1} &= \alpha k^{\alpha-1} - \theta \left(1+r_1\right)
\end{align*}
which is strictly positive given arguments already presented above. 

We can now compute the derivative $F'_{k_1}$
\begin{align} \label{eq:F_function_capital_accumulation}
    F'_{k_1} &= \alpha \times \left(\alpha-1\right) k^{\alpha-2} - \frac{\left(1-\theta\right)}{\beta} \times \mu \times \left[ \alpha k^{\alpha} - \theta \left(1+r_1\right) \right] \\
    & - \frac{\left(1-\theta\right)}{\beta}  \left[k^{\alpha}_1 - \theta k_1 \left(1+r_1\right) \right] \times \bigg\{\left(1-\theta\right) \left[k^{\alpha}_0 - b_0 \left(1+r_0\right) - k_1 \left(1-\theta \right)\right]^{-2} \\
    & - \left(-1\right) \times \beta \left(1+r_1\right) \left[ k^{\alpha}_1 - \theta k_1 \left(1+r_1\right) \right]^{-2} \times \left[ \alpha k^{\alpha-1} - \theta \left(1+r_1\right) \right] \bigg\}
\end{align}
In order to determine the sign of $F_{k_1}$ we evaluate line by line of the equation above. We begin with
\begin{align*}
    \underbrace{\alpha}_{>0} \times \underbrace{\left(\alpha-1\right) k^{\alpha-2}}_{<0} \overbrace{-\underbrace{ \frac{\left(1-\theta\right)}{\beta} \times \mu \times \left[ \alpha k^{\alpha} - \theta \left(1+r_1\right) \right]}_{>0}}^{<0} \quad <0
\end{align*}
which is strictly negative. The term in brackets is strictly positive as
\begin{align*}
     \left(1-\theta\right) \left[\underbrace{k^{\alpha}_0 - b_0 \left(1+r_0\right) - k_1 \left(1-\theta \right)}_{c_0}\right]^{-2} > 0
\end{align*}
and 
\begin{align*}
    - \left(-1\right) \times \beta \left(1+r_1\right) \left[ k^{\alpha}_1 - \theta k_1 \left(1+r_1\right) \right]^{-2} \times \left[ \alpha k^{\alpha-1} - \theta \left(1+r_1\right) \right] >0
\end{align*}
Consequently
\begin{align*}
    & - \frac{\left(1-\theta\right)}{\beta}  \left[k^{\alpha}_1 - \theta k_1 \left(1+r_1\right) \right] \times \bigg\{\left(1-\theta\right) \left[k^{\alpha}_0 - b_0 \left(1+r_0\right) - k_1 \left(1-\theta \right)\right]^{-2} \\
    & - \left(-1\right) \times \beta \left(1+r_1\right) \left[ k^{\alpha}_1 - \theta k_1 \left(1+r_1\right) \right]^{-2} \times \left[ \alpha k^{\alpha-1} - \theta \left(1+r_1\right) \right] \bigg\}
\end{align*}
is also negative. Thus, the whole term $F'_{k_1}$ is negative. Consequently, we can conclude that 
\begin{align*}
    \frac{\partial k_1 }{\partial 1+r_1} = - \frac{\overbrace{F'_{1+r_1}}^{<0}}{\underbrace{F'_{k_1}}_{<0}} < 0
\end{align*}
Note that 
\begin{align*}
    |\left(F'_{k_1}\right)^{\text{Unconstrained}}| < |\left(F'_{k_1}\right)^{\text{Constrained}}| 
\end{align*}
as $F'_{k_1}$ is negative and the additional term for constrained entrepreneur further increases its absolute value. Thus, for the case of two entrepreneurs with the same parameters $\alpha,\theta$, initial capital shock $k_0$, and close enough initial stock of debt $b_0$ such that one entrepreneur is unconstrained and a second one is constrained, the response of the optimal stock of capital $k_1$ with respect to the gross interest rate $1+r_1$ is greater in magnitude for the unconstrained entrepreneur as the derivative of the implicit function $F$ with respect to $1+r_1$, i.e. $F'_{1+r_1}$, in the numerator is smaller in magnitude for the constrained entrepreneur, and the derivative of the implicit function $F$ with respect to $k_1$, i.e. $F'_{k_1}$ is greater in magnitude for the constrained entrepreneur. This finishes the proof of Proposition 1.     $\blacksquare$

\noindent
\textbf{Proof of Proposition 2:} Next, we prove that relaxing the leverage constraint, i.e., increasing parameter $\theta$, leads to an increase in the entrepreneur's optimal capital accumulation, assuming that the entrepreneur remains constrained. The derivative can be computed as
\begin{align*}
    \frac{\partial k_1}{\partial \theta} &= - \frac{F'_{\theta}}{F'_{k_1}}
\end{align*}
Below are the terms of $F\left(k_1,1+r_1,\theta\right)$ which depend on $\theta$
\begin{align*}
   F\left(k_1,1+r_1,\theta\right) &= -\left[\frac{1}{k^{\alpha}_0 - b_0 \left(1+r_0\right) - k_1 \left(1-\theta \right)} - \frac{\beta \left(1+r_1\right)}{k^{\alpha}_1 - \theta k_1 \left(1+r_1\right)} \right] \times \\
   & \times \frac{1}{\beta} \left(1-\theta\right) \{k^{\alpha}_1 - \theta k_1 \left(1+r_1\right) \}
\end{align*}
We study each factor of the multiplication term above. The term in square brackets can be shown to be decreasing in $\theta$ 
\begin{align*}
    \underbrace{\left[\frac{1}{\underbrace{k^{\alpha}_0 - b_0 \left(1+r_0\right) - k_1 \left(1-\theta \right)}_{\text{Decreasing in $\theta$}}} - \underbrace{\frac{\beta \left(1+r_1\right)}{k^{\alpha}_1 - \theta k_1 \left(1+r_1\right)}}_{\text{Increasing in $\theta$}} \right]}_{\text{Decreasing in $\theta$}}
\end{align*}
This has intuitive sense as the squared bracket term represents the Lagrange multiplier  on the leverage constraint, i.e, $\mu$. This means that more relaxed leverage constraint leads to a lower Lagrange multiplier.

The second term, in curved brackets is clearly decreasing in $\theta$
\begin{align*}
    \{k^{\alpha}_1 - \theta k_1 \left(1+r_1\right) \}
\end{align*}
This has intuitive sense as a more relaxed leverage constraint of a constrained entrepreneur would lead them to increase its debt and thus reduce its consumption in the final period. Finally, the term in the middle,
\begin{align*}
    \frac{1-\theta}{\beta}
\end{align*}
is clearly decreasing in $\theta$. 

In order to compute the derivative of our function $F\left(k_1,1+r_1,\theta\right)$ with respect to $\theta$ we re label our terms as
\begin{align*} \small
   F\left(k_1,1+r_1,\theta\right) &= - \underbrace{\frac{\left(1-\theta\right)}{\beta} 
   \left[\frac{1}{k^{\alpha}_0 - b_0 \left(1+r_0\right) - k_1 \left(1-\theta \right)} - \frac{\beta \left(1+r_1\right)}{k^{\alpha}_1 - \theta k_1 \left(1+r_1\right)} \right]}_{\Lambda_1 \left(\theta\right)} \times  \underbrace{\{k^{\alpha}_1 - \theta k_1 \left(1+r_1\right) \}}_{\Lambda_1 \left(\theta\right)}
\end{align*} \normalsize
Thus,
\begin{align*} \small
    F_{\theta} = - \left[\frac{\partial \Lambda_0}{\partial \theta}\times \Lambda_1 \left(\theta\right)+ \frac{\partial \Lambda_1}{\partial \theta}\times \Lambda_0 \left(\theta\right)\right]
\end{align*} \footnotesize
Then
\begin{align*}
    \frac{\partial \Lambda_0}{\partial \theta} &= \frac{1-\theta}{\beta} \times \bigg\{\left(-1\right)\times \left[k^{\alpha}_0 - b_0 \left(1+r_0\right) - k_1 \left(1-\theta \right)\right]^{-2} \times k_1 \\
    & \qquad - k_1 \left(1+r_1\right) \beta \left(1+r_1\right) \left[ k^{\alpha}_1 - \theta k_1 \left(1+r_1\right)\right]^{-2}
    \bigg\} \\
    & + \left[\frac{1}{k^{\alpha}_0 - b_0 \left(1+r_0\right) - k_1 \left(1-\theta \right)} - \frac{\beta \left(1+r_1\right)}{k^{\alpha}_1 - \theta k_1 \left(1+r_1\right)} \right] \times \frac{\left(-1\right)}{\beta} < 0
\end{align*}
which makes
\begin{align*}
    \frac{\partial \Lambda_0}{\partial \theta}\times \Lambda_1 \left(\theta\right) <0
\end{align*}
as $\Lambda_1>0$ given that $c_1$ is positive given the assumptions on preferences.

\noindent 
Next, we turn to computing derivative $\partial \Lambda_1 / \partial \theta$, which is equal to
\begin{align*}
    \frac{\partial \Lambda_1}{\partial \theta} = - k_1 \left(1+r_1\right) < 0
\end{align*}
Given that $\Lambda_0>0$ as the entrepreneur is constrained, i.e. $\mu>0$, 
\begin{align*}
    \frac{\partial \Lambda_1}{\partial \theta}\times \Lambda_0 \left(\theta\right) < 0
\end{align*}
Consequently, we can conclude that
\begin{align*}
    F_{\theta} = - \overbrace{\left[\frac{\partial \Lambda_0}{\partial \theta}\times \Lambda_1 \left(\theta\right)+ \frac{\partial \Lambda_1}{\partial \theta}\times \Lambda_0 \left(\theta\right)\right]}^{<0} > 0
\end{align*}
Using the result derived above that $F'_{k_1}<0$, we can conclude that relaxing the constrained entrepreneur's leverage constraint by increasing $\theta$ leads to an increase in the optimal capital accumulation as,
\begin{align*}
    \frac{\partial k_1}{\partial \theta} &= - \frac{\overbrace{F'_{\theta}}^{>0}}{\underbrace{F'_{k_1}}_{<0}} > 0
\end{align*}
This finishes the proof of Proposition 2.     $\blacksquare$

\end{document}